\documentclass[twocolumn]{aastex61}
\usepackage{lineno}

\usepackage{graphicx}
\usepackage{subfigure}
\usepackage{multirow}
\usepackage{comment}
\usepackage{natbib}
\usepackage{hyperref}
\usepackage{mathtools}
\usepackage{longtable}
\usepackage{mathrsfs}
\usepackage{fontenc}
\usepackage{color}
\usepackage{url}
\usepackage{bm}
\usepackage{hyperref}
\usepackage{gensymb}
\usepackage{pifont}
\newcommand{\sersic}{{S\'{e}rsic}}

\newcommand{\mum}{$\mu$m}

\newcommand{\ciii}{C\,{\sc iii}\rm]}

\newcommand{\hi}{H\,{\sc i}\rm}

\newcommand{\hei}{He\,{\sc i}\rm}
\newcommand{\heii}{He\,{\sc ii}\rm}

\newcommand{\angs}{~\AA}

\newcommand{\lya}{\mbox{Ly$\alpha$}}

\newcommand{\civ}{\mbox{C\,{\sc iv}}}
\newcommand{\mgii}{\mbox{Mg\,{\sc ii}}}

\newcommand{\feii}{\mbox Fe\,{\sc ii}}
\newcommand{\ha}{\mbox{H$\alpha$}}
\newcommand{\hb}{\mbox{H$\beta$}}
\newcommand{\hg}{\mbox{H$\gamma$}}

\newcommand{\sii}{\mbox{[S\,{\sc ii}]}}

\newcommand{\nii}{[N\,{\sc ii}] }

\newcommand{\oiii}{[O\,{\sc iii}]}

\newcommand{\oi}{[O\,{\sc i}] }

\submitjournal{ApJ}

\begin{document}


\title{The Host Galaxy and Rapidly Evolving Broad-line Region in the Changing-look Active Galactic Nucleus 1ES\,1927+654}

\shorttitle{The Host Galaxy and BLR of 1ES\,1927+654}

\shortauthors{Li et al.}

\author{Ruancun Li}
\affil{Kavli Institute for Astronomy and Astrophysics, Peking University, Beijing 100871, China}
\affil{Department of Astronomy, School of Physics, Peking University, Beijing 100871, China}

\author{Luis C. Ho}
\affil{Kavli Institute for Astronomy and Astrophysics, Peking University, Beijing 100871, China}
\affil{Department of Astronomy, School of Physics, Peking University, Beijing 100871, China}

\author{Claudio Ricci}
\affil{N\'ucleo de Astronom\'\i ıa de la Facultad de Ingenier\'\i a, Universidad Diego Portales, Av. Ej\'ercito Libertador 441, Santiago 22, Chile}
\affil{Kavli Institute for Astronomy and Astrophysics, Peking University, Beijing 100871, China}
\affil{George Mason University, Department of Physics \& Astronomy, MS 3F3, 4400 University Drive, Fairfax, VA 22030, USA}

\author{Benny Trakhtenbrot}
\affil{School of Physics and Astronomy, Tel Aviv University, Tel Aviv 69978, Israel}

\author{Iair Arcavi}
\affil{School of Physics and Astronomy, Tel Aviv University, Tel Aviv 69978, Israel}
\affil{CIFAR Azrieli Global Scholars program, CIFAR, Toronto, Canada}

\author{Erin Kara}
\affil{MIT Kavli Institute for Astrophysics and Space Research, 70 Vassar Street, Cambridge, MA 02139, USA}

\author{Daichi Hiramatsu}
\affiliation{Center for Astrophysics \textbar{} Harvard \& Smithsonian, 60 Garden Street, Cambridge, MA 02138-1516, USA}
\affiliation{The NSF AI Institute for Artificial Intelligence and Fundamental Interactions}

\begin{abstract}
Changing-look active galactic nuclei (AGNs) present an important laboratory to understand the origin and physical properties of the broad-line region (BLR). We investigate follow-up optical spectroscopy spanning $\sim 500$ days after the outburst of the changing-look AGN 1ES\,1927+654. The emission lines displayed dramatic, systematic variations in intensity, velocity width, velocity shift, and symmetry. Analysis of optical spectra and multi-band images indicate that the host galaxy contains a pseudobulge and a total stellar mass of $3.56_{-0.35}^{+0.38} \times 10^{9}\, M_\odot$.  Enhanced continuum radiation from the outburst produced an accretion disk wind, which condensed into BLR clouds in the region above and below the temporary eccentric disk. Broad Balmer lines emerged $\sim 100$ days after the outburst, together with an unexpected, additional component of narrow-line emission. The newly formed BLR clouds then traveled along a similar eccentric orbit ($e \approx 0.6$). The Balmer decrement of the BLR increased by a factor of $\sim 4-5$ as a result of secular changes in cloud density. The drop in density at late times allowed the production of \hei\ and \heii\ emission. The mass of the black hole cannot be derived from the broad emission lines because the BLR is not virialized. Instead, we use the stellar properties of the host galaxy to estimate $M_\mathrm{BH} = 1.38_{-0.66}^{+1.25} \times 10^{6}\, M_\odot$.  The nucleus reached near or above its Eddington limit during the peak of the outburst. We discuss the nature of the changing-look AGN 1ES\,1927+654 in the context of other tidal disruption events.
\end{abstract}


\section{Introduction}

Broad emission lines are a hallmark characteristic of type~1 active galactic nuclei (AGNs).  Arising from rapidly moving, dense photoionized gas near the supermassive black hole (BH), the emission lines from the broad-line region (BLR) are illuminated by the ultraviolet (UV) continuum from the accretion disk (e.g., \citealp{Davidson1979RvMP,Kwan1981ApJ}).  The size of the BLR ($R_\mathrm{BLR}$), which can be estimated by cross correlating the light curve of the continuum with that of the broad lines, scales with the AGN luminosity ($R_\mathrm{BLR}-L$ correlation; e.g., \citealp{Kaspi2000ApJ,Peterson2004ApJ,Bentz2013ApJ}).  In the classical picture, the BLR consists of a collection of clouds virialized by the gravitational potential of the central BH.  This is supported by the discovery that the full width at half maximum (FWHM) of broad \hb\ is proportional to the radius of the BLR: FWHM$_\mathrm{H\beta} \propto R_\mathrm{BLR}^{-1/2}$ (e.g., \citealp{Peterson1999, Wang2020ApJ}).  Recently, a generalized BLR model was developed based on the combination of Keplerian rotation and  radial motion of the clouds \citep{Pancoast2014MNRAS,Li2018ApJ}. This model has been used to study the geometry of the BLR using spatially resolved interferometric spectra \citep{Gravity2018Natur,Gravity2020AA} and velocity-resolved reverberation mapping (RM) observations \citep{Grier2017ApJ,Williams2018ApJ}. The dynamics of the BLR may be a combination of rotation and outflow in a way that depends on the properties of individual objects (e.g., \citealp{Du2016aApJ,Horne2021ApJ}).

A subset of AGNs display transient behavior in which their broad emission lines can appear or disappear, accompanied by large-amplitude continuum variability, switching between type~2 and type~1 optical classifications. Several dozens of these so-called changing-look AGNs have been discovered over the past decade (e.g., \citealp{Denney2014ApJ,Shappee2014ApJ,LaMassa2015ApJ,Rumbaugh2018ApJ,MacLeod2019ApJ,Trakhtenbrot2019ApJ,Yang2019ApJ}).  Distinct from AGNs that exhibit dramatic X-ray variability due to varying line-of-sight column density (e.g., \citealp{Matt2003MNRAS,Risaliti2006ESASP,Rivers2015ApJ,Ricci2016ApJ}), optical changing-look transitions might be related to dramatic changes of the accretion flow, similar to those commonly observed in outbursting X-ray binaries \citep{Homan2005ApSS,Remillard2006ARAA,Done2007AARv}.  Observations (e.g., \citealp{Ho2008ARAA,Ho2009aApJ,Noda2018MNRAS,Liu2020MNRAS}) indicate that AGN accretion flows undergo changes in structure and radiative efficiency when the mass accretion rate ($\dot{M}$) reaches certain critical values relative to the Eddington accretion rate ($\dot{M}_\mathrm{E}$): whereas the accretion flow is geometrically thin and radiatively efficient when $\dot{M} \simeq 0.1-1\dot{M}_\mathrm{E}$ (standard thin disk; \citealp{Shakura1973AA,Novikov1973blho}), it becomes geometrically thick and radiatively inefficient when $\dot{M} \gtrsim \dot{M}_\mathrm{E}$ (slim disk; \citealp{Katz1977ApJ,Abramowicz1988ApJ,Sadowski2009ApJS}) or $\dot{M} \lesssim 0.01\dot{M}_\mathrm{E}$ (advection-dominated accretion flow; \citealp{Ichimaru1977ApJ,Narayan1994ApJ}). Consequent changes of both the reprocessing power and covering factor of the BLR result in the appearance and disappearance of broad emission lines \citep{Dexter2019ApJ}. Large-amplitude variability of changing-look AGNs in the mid-infrared (IR) further supports this picture \citep{Sheng2017ApJ}.

Alternatively, nuclear tidal disruption events (TDEs; for a review, see \citealp{Gezari2021ARAA}) could also explain the sudden emergence of broad emission lines and the blue continuum (e.g., \citealp{Merloni2015MNRAS,Trakhtenbrot2019ApJ,Ricci2020ApJL}). The central supermassive BH can accrete a fraction of the mass of the tidally disruped star, causing a flare that peaks in the extreme-UV band and that can extend to the optical and X-rays (e.g., \citealp{Rees1988Nature,Saxton2019AN}). TDEs can also generate highly ionized outflows, detectable as blueshifted broad hydrogen Balmer or helium lines \citep{Miller2015Nature,Hung2019ApJ}, as well as P~Cygni-like absorption features at X-ray energies \citep{Kara2018MNRAS}.  However, the mechanisms that produce broad emission lines in TDEs are still controversial. Observationally, broad  ($1-2\times 10^4\, \rm km \, s^{-1}$) H and He lines usually dominate the spectra of optically selected TDEs \citep{Arcavi2014ApJ}. Generally observed in emission, optical lines can have complex and asymmetric shapes (e.g., \citealp{Gezari2012Nature,Arcavi2014ApJ}). The width of the emission lines typically decreases with time while the continuum luminosity drops \citep{Holoien2016aMNRAS,Holoien2016bMNRAS,Leloudas2019ApJ}, opposite of what is expected from the virialized BLR of type~1 AGNs. While the unbound material from the disrupted star initially was considered to be the primary contributor to the broad emission lines \citep{Kasen2010ApJ,Clausen2011ApJ}, later hydrodynamical simulations suggest that the debris stream is confined to a negligible surface area and does not contribute significantly to either the continuum or line emission \citep{Guillochon2014ApJ}. Instead, broad emission lines may be produced in the region above and below the elliptical accretion disk. \citet{Liu2017MNRAS} suggest that the optical emission lines are associated with the accretion flow, such that relativistic broadening can account for the double-peak broad \ha\ profile observed in some sources (e.g., \citealp{Arcavi2014ApJ}).

Both scenarios suggest that the accretion rate plays a role in determining the AGN type, since the optical continuum becomes bluer and brighter when CL AGNs turn on, and vice versa (e.g., \citealp{Yang2018ApJ}). It is interesting that the type~1 phase of CL AGNs lasts only $\sim 10$\,yr (e.g., \citealp{Denney2014ApJ,McElroy2016AA}). The variable accretion rate scenario requires that the AGN hovers near the state transition threshold of $\dot{M} \approx 0.01$), beyond which the ionizing luminosity significantly changes. Indeed, phase-transition CL AGNs usually have bolometric luminosities of a few percent of $L_\mathrm{Edd}$ (e.g., \citealp{Noda2018MNRAS}). By contrast, the scenario involving TDEs, which can have a variety of possible penetration factors and involve diverse properties of the disrupted star, does not select BHs with a preferential $\dot{M}$. And while the flux can vary over a large dynamical range, the time-dependent fallback accretion rate should follow a universal for of $\dot{M} \propto t^{-5/3}$ (\citealp{Rees1988Nature,Lodato2011MNRAS}). Therefore, detailed monitoring of the evolution of the BH accretion rate can, in principle, distinguish between these two scenarios.

A changing-look event was recently reported and studied \citep{Trakhtenbrot2019ApJ,Ricci2020ApJL,Ricci2021ApJS} in the nearby ($z=0.019422$) galaxy 1ES\,1927+654, a previously known type~2 (narrow-line) AGN \citep{Boller2003AA,Tran2011ApJ}.  Following its discovery by the All-Sky Automated Survey for Supernovae (ASAS-SN; \citealp{Shappee2014ApJ}), its $V$-band flux increased by at least 2 magnitudes in March 2018 (\citealp{Nicholls2018ATel}). Subsequent optical spectroscopic follow-up spanning $\sim 500$ days found the emergence of broad Balmer emission several weeks after the outburst, followed by broad \lya\ emission detected on 28 August 2018 \citep{Trakhtenbrot2019ApJ}.  This is the first case of a source whose changing-look transition has been observed. 

This work studies in detail the properties of the optical lines of 1ES\,1927+654, in the context of the dramatic variations of the BLR. We also constrain the mass of the BH through the properties of the host galaxy. We demonstrate that the virial mass of the BH calculated through the traditional method of assuming a virialized BLR is inconsistent with other independent estimates.  Section~\ref{sec:sec2} describes our photometric observations and analysis.  Section~\ref{sec:sect3} illustrates our optical spectral fitting and discusses the properties of the host galaxy and emisssion lines. We then summarize the spectral evolution of 1ES\,1927+654 after the optical outburst, and estimate the BH mass in Section~\ref{sec:specevol}. Section~\ref{sec:discussion} offers a proposed physical picture of the changing-look process and evolving BLR. Conclusions are given in Section~\ref{sec:summary}. For the adopted $\Lambda$CDM cosmology ($\Omega_\mathrm{m} = 0.308$, $\Omega_\Lambda = 0.692$, and $H_0 = 67.8\, \rm \; km \, s^{-1} \, Mpc^{-1}$; \citealp{Planck2016AA}), the luminosity distance of 1ES\,1927+654 is $87.2$ Mpc.

\section{Photometric Analysis}
\label{sec:sec2}

The flux of the source changed significantly in the optical and UV bands following the changing-look event. \citet{Trakhtenbrot2019ApJ} provide fixed-aperture optical-to-UV photometry and light curves covering $\sim 550$ days, starting from about 50 days before the outburst. To minimize contamination from foreground stars, in this study we perform imaging decomposition (Section~\ref{sec:GALFIT}) of the four observations acquired with the Optical/UV Monitor Telescope (OM; \citealp{Mason2001AA}) onboard XMM-Newton \citep{Jansen2001AA}. These optical/UV images are crucial for estimating the bulge-to-total light ratio ($B/T$) and the stellar population of the host galaxy (Section~\ref{sec:beforeout}). To constrain the total spectral energy distribution (SED) before the outburst (Section~\ref{sec:beforeout}), we also follow a similar method to analyze the IR images from the Two-Micron All Sky Survey (2MASS; \citealp{Skrutskie2006AJ}) and the Wide-field Infrared Survey Explorer (WISE; \citealp{Wright2010AJ}).

\vfill

\subsection{Observations}
\label{sec:photoOBs}

Table~\ref{tab:photometry} summarizes the photometric data used in this study.  We use four XMM-Newton OM observations carried out between 2011 and 2019. For each observation, we extract the original data files (ODF). We used the {\tt omchain} package in Science Operations Centre SAS v18.0.0 (Science Analysis System) for image processing, after which {\tt .SIMAGE} files were generated for photometric analysis.  A maximum of six filters are available (UVW2, UVM2, UVW1, $U$, $B$, $V$), but some epochs did not cover all of them.  Near-IR ($J$, $H$, $K_s$) images acquired in 22 May 1999 were downloaded from the 2MASS archives. We cut the size of the images to 10\arcmin\ $\times$ 10\arcmin\ to ensure that there is enough area for proper sky measurement.  A similar size is used for the mid-IR W1$-$W4 images downloaded from ALLWISE \citep{Cutri2014yCat}.  A total of 68 observations of 1ES\,1927+654 are available.  We concentrate on those taken in June 2010 and December 2010, which showed less than 1 magnitude variation.

\subsection{Sky Subtraction and Masking}
\label{sec:ap}

The angular diameter of 1ES\,1927+654 is 26\farcs90, measured at the isophotal surface brightness level of 20 $K_s$ mag arcsec$^{-2}$ (2MASS Extended Source Catalog; \citealp{jarrett20002mass}). Three bright stars are located near our target of interest: star 1 ($V = 13.67$ mag), 11\farcs14 to the southwest; star 2 ($V = 15.32$ mag), 13\farcs82 to the southeast; and star 3 ($V = 14.98$ mag), 22\farcs97 to the southwest. The three stars, spectroscopically identified as G or K type \citep{Boller2003AA}, are blended with 1ES\,1927+654 in all the images used here, which have a spatial resolution of ${\rm FWHM} > 1\farcs5$.  We use {\tt GALFIT 3.0} to perform two-dimensional (2-D) imaging decomposition for proper deblending and photometric measurement (Section~\ref{sec:GALFIT}).  An accurate  estimation of the background, which {\tt GALFIT} assumes to be uniform, is needed for the analysis. However, large-scale variations may be present in the background of real images. For instance, in the near-IR bands, particularly for the 2MASS $K_s$ band, ``airglow" can produce a $\sim$200\arcsec\ gradient in the background \citep{jarrett20002mass}. For the XMM-Newton OM, ghost images from light scattered within the detector may be important\footnote{XMM-Newton Users Handbook: {\url{https://xmm-tools.cosmos.esa.int/external/xmm_user_support/documentation/uhb/omlimits.html}}}. 

We generate segmentation images following standard methods of source detection (e.g., {\tt SExtractor}), using ``sigma clipping" to estimate the root-mean-square (RMS) of the background, and then adopt 3 times the sky RMS as the threshold for source detection.  We fit a 2-D Gaussian function to convert the image segments into elliptical masks, which have the same second-order central moment as the sources. The image segments usually can be enclosed using 3 times the standard deviation of the 2-D Gaussian function in the semi-major and semi-minor axes. To ensure that all the emission is properly captured by the image segments, we enlarge the elliptical mask by extending its size by a factor of 2. The masks occupy more than $80\%$ of the sky pixels in the 10\arcmin\ $\times$ 10\arcmin\ field. To model the potential large-scale background gradient, we adopt a third-order polynomial function, which is quite efficient without overfitting the faint structures of extended galaxies \citep{jarrett20002mass}.  We subtract the best-fit background model from the original image and use it as input for {\tt GALFIT}.  Sources that are not blended with 1ES\,1927+654 are masked. We use {\tt SExtractor} to perform preliminary source deblending, using a combination of multi-thresholding and watershed segmentation \citep{Beucher1993MMIP}. We generate the mask using the same method as that used for sky subtraction, except that we do not mask our target and the three bright foreground stars.

\startlongtable
\begin{deluxetable*}{clccrccc}
\tablecaption{Summary of Photometric Analysis \label{tab:photometry}}
\tabletypesize{\scriptsize}
\tablehead{
      \colhead{Instrument}    &
      \colhead{Obs. Date}   &
      \colhead{Band} &
      \colhead{FWHM}   &
      \colhead{Magnitude} & 
      \colhead{$B/T$ } &
      \colhead{$f_\mathrm{AGN}$} &
      \colhead{$\chi^2_\nu$} \\
      \colhead{(1)} &
\colhead{(2)} &
\colhead{(3)} &
\colhead{(4)} &
\colhead{(5)} &
\colhead{(6)} &
\colhead{(7)} &
\colhead{(8)}
}
\startdata
      2MASS       & 22-05-1999 &     $J$ & $3\farcs11$ &  $13.99\pm0.11$ & $0.48\pm0.06$  & \nodata & 0.656\\
      2MASS       & 22-05-1999 &     $H$ & $3\farcs19$ &  $13.41\pm0.35$ & $0.35\pm0.24$  & \nodata & 0.752\\
      2MASS       & 22-05-1999 &   $K_s$ & $3\farcs27$ &  $12.86\pm0.11$ & $0.44\pm0.06$  & \nodata & 0.943\\
      WISE        & 27-06-2010 &     W1  & $6\farcs1 $ &  $13.43\pm0.55$ & $0.87\pm0.26$  & \nodata & 0.372\\
      WISE        & 27-06-2010 &     W2  & $6\farcs8 $ &  $12.80\pm0.50$ & $0.65\pm0.21$  & \nodata & 0.253\\
      WISE        & 27-06-2010 &     W3  & $7\farcs4 $ &  $10.42\pm0.57$ & \nodata        & \nodata & 0.277\\
      WISE        & 27-06-2010 &     W4  & $12\farcs0$ &   $8.79\pm0.63$ & \nodata        & \nodata & 0.386\\
    XMM-Newton OM & 20-05-2011\tablenotemark{a} &  UVM2 & $1\farcs96$ &  $18.62\pm0.29$ & \nodata  & \nodata &  1.201 \\ 
    XMM-Newton OM & 20-05-2011 &   UVW1  & $2\farcs14$ &  $17.46\pm0.46$ & $0.24\pm0.06$  & \nodata &  1.185 \\ 
    XMM-Newton OM & 20-05-2011 &     $V$ & $1\farcs51$ &  $16.05\pm0.14$ & $0.30\pm0.09$  & \nodata & 4.184 \\ 
    XMM-Newton OM & 05-06-2018 &   UVW2  & $2\farcs17$ &  $16.45\pm0.21$ & \nodata       & $>0.90$  & 2.947 \\
    XMM-Newton OM & 05-06-2018 &   UVM2  & $2\farcs01$ &  $16.39\pm0.11$ & \nodata       & $>0.90$  & 2.032 \\
    XMM-Newton OM & 05-06-2018 &   UVW1  & $2\farcs49$ &  $16.25\pm0.16$ & \nodata       & $>0.90$  & 3.816 \\
    XMM-Newton OM & 05-06-2018 &     $U$ & $2\farcs25$ &  $16.08\pm0.44$ & \nodata       & $>0.90$  & 8.262 \\
    XMM-Newton OM & 05-06-2018 &     $B$ & $2\farcs21$ &  $15.99\pm0.27$ & \nodata       & $>0.90$  & 12.35 \\
    XMM-Newton OM & 05-06-2018 &     $V$ & $1\farcs71$ &  $15.41\pm0.15$ & \nodata       & $0.46\pm0.06$  & 3.189 \\
    XMM-Newton OM & 07-05-2019 &   UVW2  & $2\farcs26$ &  $17.58\pm0.24$ & \nodata       & $>0.90$  & 9.314 \\
    XMM-Newton OM & 07-05-2019 &   UVM2  & $2\farcs07$ &  $17.46\pm0.15$ & \nodata       & $>0.90$  & 7.526 \\
    XMM-Newton OM & 07-05-2019 &   UVW1  & $2\farcs58$ &  $17.17\pm0.13$ & \nodata       & $>0.90$  & 10.28 \\
    XMM-Newton OM & 07-05-2019 &     $U$ & $2\farcs34$ &  $16.90\pm0.20$ & \nodata       & $0.81\pm0.08$  & 11.87 \\
    XMM-Newton OM & 07-05-2019 &     $B$ & $2\farcs01$ &  $16.56\pm0.25$ & \nodata       & $0.64\pm0.07$  & 20.60 \\
     \enddata
\tablecomments{Col. (1): Instrument. Col. (2): Date of observations. Col. (3): Filter. Col. (4): FWHM of effective PSF, generated from field stars. Col. (5): Integrated magnitude of all the components. Col. (6): Bulge-to-total ratio. Col. (7): Flux fraction of the AGN component. Col. (8): Reduced $\chi^2$ of {\tt GALFIT} model.  
\tablenotetext{a}{For the XMM-Newton OM images, the exposure was for May 2011 was 1.4 ks, while for June 2018 it was 4.5 ks and for May 2019 it was 4.4 ks.}
      }
       \end{deluxetable*}

\subsection{Imaging Decomposition}
\label{sec:GALFIT}

\begin{figure*}
    \centering
    \includegraphics[width=\textwidth]{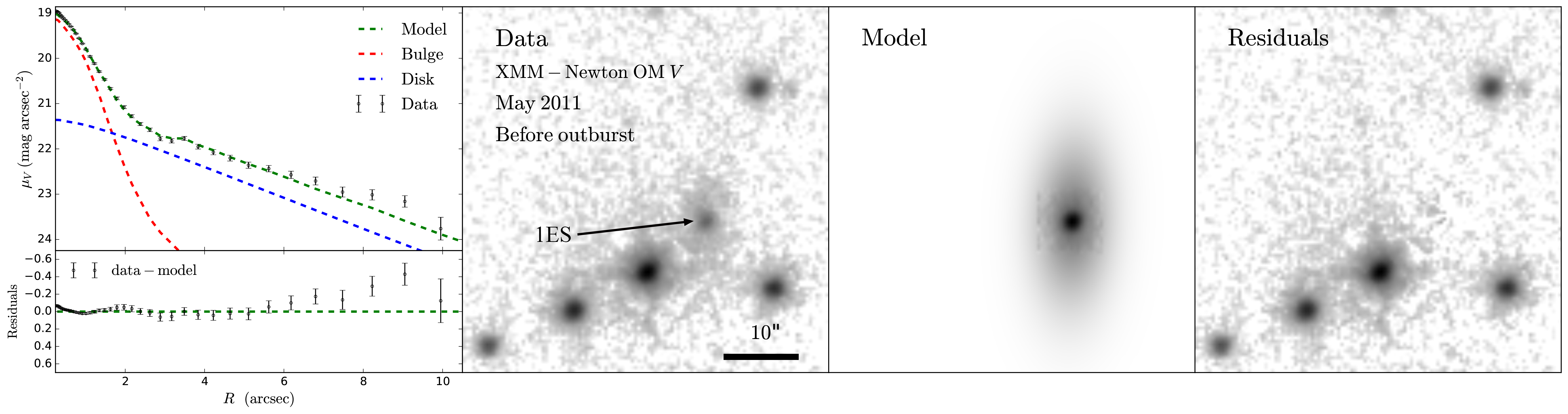}
    \includegraphics[width=\textwidth]{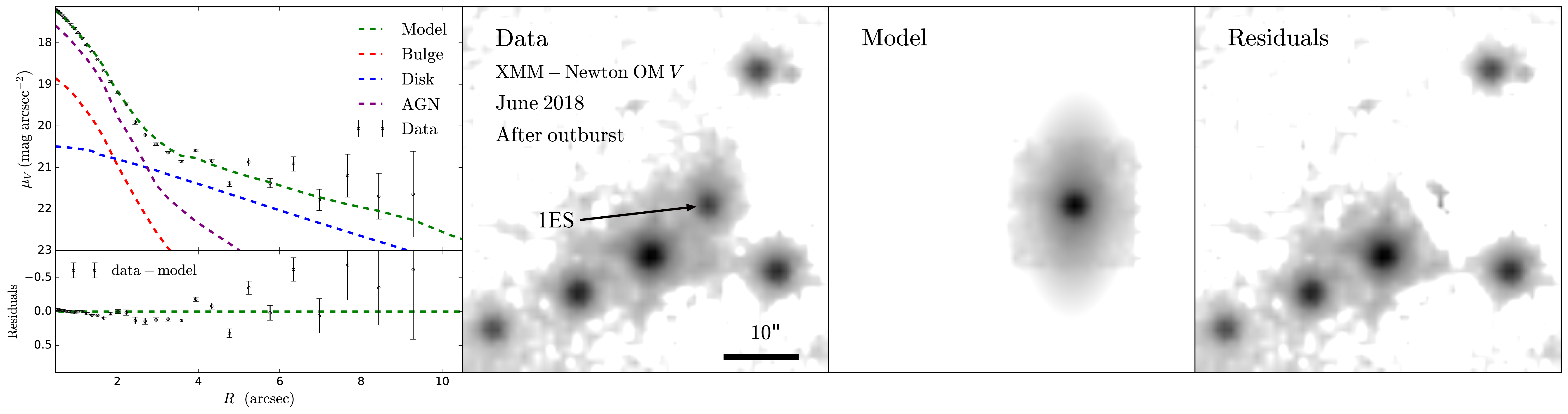}
    \caption{Example multi-component fitting of XMM-Newton OM $V$-band images. Before the outburst (first row, May 2011 image), we modeled the galaxy with a \sersic\ profile (bulge; red dashed curve in the first panel) plus an exponential profile (disk; blue dashed curve in the first panel). The overall model (green dashed curve in the first panel) shows good consistency with the data (black error bars).  After the outburst (second row, June 2018 image), we add a point source to account for the emission from the AGN (purple dashed curve).  The second column shows the observed image; the third column shows the {\tt GALFIT} model; and the fourth column shows the residuals.}
    \label{fig:GALFIT}
\end{figure*}

In light of the severe blending by foreground stars, it is a challenge to obtain reliable photometry for our target of interest.  We simultaneously deblend 1ES\,1927+654 and its three nearby stars using {\tt GALFIT}, considering a 55\arcsec\ box centered on our target, which is large enough to enclose all bright foreground stars while avoiding contamination from other nearby sources (Figure\,\ref{fig:GALFIT}).  We use the {\tt EPSFBuilder} task from the Python package {\tt photutils} to build the point-spread function (PSF) of each image by fitting bright, unsaturated, isolated stars in the field.  We adopt a convolution box of size $40\arcsec \times40 \arcsec$, which is roughly 20 times the FWHM of the PSF and is large enough to cover the wings of the PSF. Sigma images were made directly from the original data, based on the Poisson noise of each pixel, which is the quadrature sum of the contribution from the source and the local sky background \citep{Peng2010AJ}.

We start our fitting with the May 2011 XMM-Newton OM $V$-band image, not only because it has the highest resolution (PSF $\mathrm{FWHM}=1\farcs51$), but also because the $V$-band emission at this epoch was dominated by the host galaxy.  The fit takes into account several physical considerations: (1) because of the galaxy's relatively low stellar mass ($M_\ast = 3.56\times 10^9\, M_\odot$; Section~\ref{sec:beforeout}), we do not expect a S\'ersic index higher than 6 \citep{Gao2020ApJS}; (2) the axis ratio ($b/a$) should be higher than 0.15, since only $1\%$ of galaxies with $M_\ast \simeq 10^9\,M_\odot$ have $b/a < 0.15$ \citep{Sanchez2010MNRAS}; and (3) different components of the same objects should have the same central position.  The best-fitting result suggests that the galaxy can be described by the sum of a \sersic\ component with index $n=1.5$ (effective radius $r_e=0\farcs42$, axis ratio $b/a=0.24$, position angle ${\rm PA} = 75.7\degree$) plus an exponential profile ($r_e=6\farcs83$, $b/a=0.44$, ${\rm PA} = 87.9\degree$). These two components are clearly evident in the one-dimensional (1-D) surface brightness profile (the first column of Figure\,\ref{fig:GALFIT}) generated using the {\tt IRAF} task {\tt ellipse}. The compact (red dashed curve) and extended (blue dashed curve) components can be interpreted as the bulge and the disk of the host galaxy, respectively. The uncertainty of the integrated magnitude of each component follows

\begin{equation}
    \sigma_{m}^2 = \sigma_{\rm stat}^2 + \sigma_{\rm syst}^2,
    \label{equ:magsigma}
\end{equation}

\noindent
where $\sigma_{\rm stat}$ represents the statistical uncertainty, which is analytically given by {\tt GALFIT} based on the covariance matrix of the parameters, and $\sigma_{\rm syst}$ is the systematic uncertainty of our image decomposition method, which we estimate from the standard deviation of the integrated magnitudes of different bulge models with \sersic\ indexes fixed from $n=1$ to $n=5$.

Table~\ref{tab:photometry} summarizes the estimates of $B/T$ that we deem to be reliable, in bands in which stellar emission dominates. As expected, the bulge tends be more prominent at longer wavelengths. With $B/T = 0.44$ in the $K_s$ band, the bulge of 1ES\,1927+654 is about twice as dominant as other galaxies of similar mass.  For example, galaxies with $M_\ast \simeq 10^9\, M_\odot$ typically have $B/T \lesssim 0.2$ \citep{Moffett2016MNRAS}.  The host galaxy became severely contaminated by the AGN after the outburst that triggered the changing-look event.  Consequently, we add a point source component to represent the AGN contribution for the 2018 and 2019 observations.  When modeling the post-outburst images, the parameters of the host galaxy were fixed to the best-fit values before the outburst, and we only allowed the normalization to adjust.  We calculate the fraction of AGN emission ($f_\mathrm{AGN}$) by dividing the flux of the point source component by the total flux, and we estimate its uncertainties following Equation \ref{equ:magsigma}. For observations in which the central AGN dominates the total emission, the host galaxy component cannot be measured with confidence, and we simply provide a lower limit for $f_\mathrm{AGN}$.

\section{Spectroscopic Analysis}
\label{sec:sect3}

The optical spectrum of 1ES\,1927+654 acquired by \citet{Boller2003AA} 16 years prior to the outburst (23 December 2017) shows prominent narrow emission lines superposed on a starlight-dominated continuum, and no evidence of either Fe~{\sc ii} multiplets or broad lines. With \oiii~$\lambda$5007/\hb\ = 14.6 and \nii~$\lambda$6584/\ha\ = 0.6, the spectrum is typical of that of Seyfert~2 galaxies \citep{Veilleux1987ApJS,Ho1993ApJ}.  The 26 post-outburst spectra presented by \citet{Trakhtenbrot2019ApJ} first reveal a blue continuum and, about 3 months later, broad emission lines. Together with eight additional spectra acquired after \citet{Trakhtenbrot2019ApJ}, here we uniformly analyze all 34 spectra using multi-component decomposition, separately focusing on the global spectral modeling before and after the outburst, the evolution of the broad emission lines, and their physical interpretation.

\subsection{Flux Calibration}
\label{sec:fluxcal}

As the spectra were taken over the course of many observing runs, under a variety of conditions and using diverse instrument configurations, they must be homogenized onto a common flux scale prior to further analysis.  The spectra presented in \cite{Trakhtenbrot2019ApJ} were originally scaled to match the $o$-band photometry of the Asteroid Terrestrial-impact Last Alert System (ATLAS; \citealp{Tonry2018PASP}). However, because the effective wavelength of the $o$ band ($\sim6800$\angs) is contaminated by \ha, which was very bright $\sim 100-200$ days after the outburst when a broad component surfaced, here we seek a different strategy for flux calibration.  We use, instead, the flux of the \oiii~$\lambda 5007$ line measured prior to the outburst as the reference for scaling the flux of the new observations, under the conventional assumption (but see Section~\ref{sec:narrow_var}) that the narrow-line region has remained constant over the monitoring period. Given the pre-outburst \oiii\ luminosity of $2.6 \times 10^{40}\rm\,erg\,s^{-1}$ \citet{Boller2003AA}, the narrow-line region subtends a radius of $\sim 1\,\rm kpc$ according to the size-luminosity relation of \cite{Chen2019}, a scale that is well captured by our later spectroscopic observations (slit widths $\sim 1\farcs5-2\arcsec$; \citealt{Trakhtenbrot2019ApJ}).

\startlongtable
\begin{deluxetable}{lcr}
\tablecaption{Best-fit Parameters before the Outburst\label{tab:host}}
\tabletypesize{\scriptsize}
\tablehead{
\colhead{Component}             &
\colhead{Parameter}   &
\colhead{Best-fit Value} \\
\colhead{(1)} &
\colhead{(2)} &
\colhead{(3)}
}
\startdata
Reddening & $A_V$ &   $0.42^{+0.21}_{-0.08}$ \\
\hline
SSP1 & log $M_\ast$ [$M_\odot$] &  $8.17^{+0.28}_{-0.10}$ \\
     & $t$ [Gyr] & $0.06^{+0.02}_{-0.01}$ \\
SSP2 & log $M_\ast$ [$M_\odot$] &  $9.05^{+0.02}_{-0.05}$ \\
     & $t$ [Gyr] & $0.98^{+0.02}_{-0.10}$ \\
\hline
Power law & $\alpha_\nu$ & $0.71^{+0.91}_{-0.90}$ \\
          & log $L_{5100}$ [$\rm erg\,s^{-1}$] & $41.47^{+0.31}_{-0.41}$\\ 
\hline
Torus & $a$ & $-0.52^{+0.34}_{-0.28}$ \\
      & $h$ & $0.23^{+0.20}_{-0.16}$ \\
      & $N_0$ &  $9.97^{+0.70}_{-1.94}$ \\
      & $i$ &  $13.82^{+16.32}_{-9.05}$ \\
      & log $L_\mathrm{torus}$ [$\rm erg\,s^{-1}$] & $40.34^{+0.06}_{-0.06}$\\ 
\enddata
\tablecomments{Best-fit parameters of the optical to MIR SED before the outburst (see Section \S\ref{sec:beforeout}). Col. (1): Component of the overall model. Col. (2): Parameter of the model and its units. Col. (3): Best-fit value of the parameter. Each SSP is described by its stellar mass ($M_\ast$) and age ($t$). For the power-law AGN component, we allow the slope ($\alpha_\nu$) and normalization ($L_{5100}$) to vary. The torus component is described by five free parameters: (1) power-law index of the radial dust cloud distribution ($a$); (2) dimensionless scale height ($h \equiv H/r$); (3) number of clouds along an equatorial line-of-sight ($N_0$); (4) inclination angle ($i$); and (5) integrated luminosity ($L_\mathrm{torus}$).
}
\end{deluxetable}

\vfill

\subsection{Before the Outburst}
\label{sec:beforeout}

The pre-outburst spectrum \citep{Boller2003AA} taken in June 2001 covers $4000-7000$\angs\ with a spectral resolution of 6\angs\ (3 pixels) and a signal-to-noise ratio (SNR) $\sim 30$.  We first scale the absolute flux of the spectrum to match our photometric measurements from the 2011 XMM-Newton OM observation (factor $13.4\pm2.7$; Section~\ref{sec:photoOBs}), since at this time the optical continuum was dominated by the host galaxy and variability should be negligible on a timescale of $\sim 10$ years.  We then correct the spectrum for a Galactic dust extinction of $A_V=0.23$ mag \citep{Schlafly2011ApJ} using the extinction curve of \citet{Cardelli1989ApJ} and convert to the rest frame assuming a redshift of $z=0.01942$, which was calculated from the narrow emission lines by \cite{Trakhtenbrot2019ApJ}.

\begin{figure*}
\centering
\includegraphics[width=0.8\textwidth]{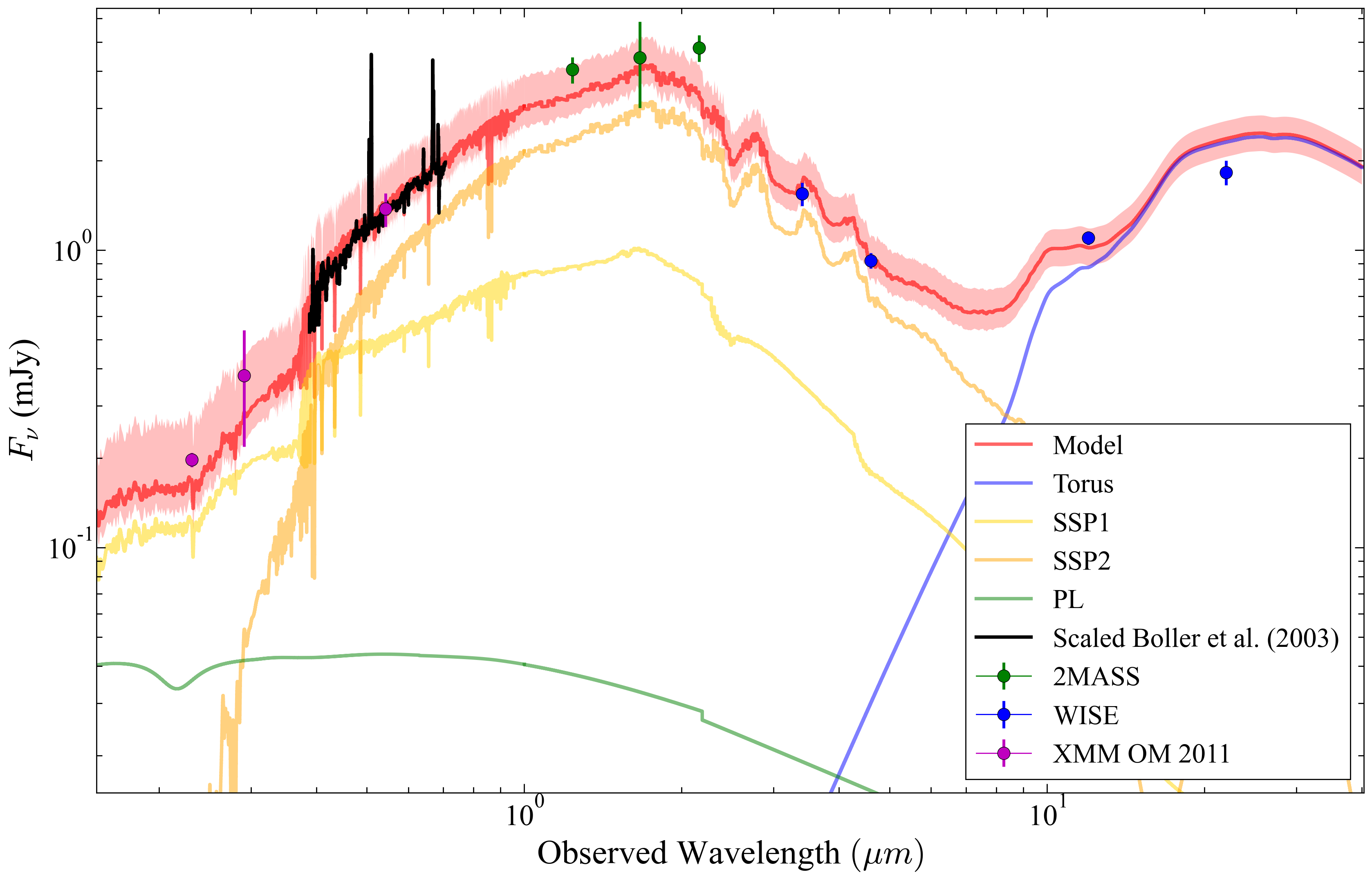}
\caption{Optical to mid-IR SED before the outburst (Section~\ref{sec:beforeout}). The black curve shows the June 2001 optical spectrum \citep{Boller2003AA} scaled to match the 2011 XMM-Newton OM $V$-band photometry in the observed frame. Error bars show our photometric mesurements of the 2011 OM images in the UVM2, UVW1, and $V$ bands (purple), 2MASS images in the $J$, $H$, and $K_s$ bands (green), and WISE images in the W1 to W4 bands (blue). The red curve and the red shaded region show our best-fit model and its $1\,\sigma$ uncertainty, respectively, which consists of two SSPs (\citealp{Bruzual2003MNRAS}; golden and orange curve), a featureless power-law component (green curve), and a dusty torus (\citealp{Honig2017ApJ}; blue curve). }
 \label{fig:hostga}
\end{figure*}

We incorporate the optical spectrum into a global fit of the broad-band SED (Figure\,\ref{fig:hostga}), covering $\sim 2300$ \AA\ to 22.2 $\mu$m, using the Bayesian Markov chain Monte Carlo (MCMC) inference method developed by \cite{Shangguan2018ApJ}, with the primary aim of estimating the total stellar mass of the galaxy. After masking the emission lines\footnote{Rest-frame wavelength intervals: 3820$-$3900, 4060$-$4140, 4300$-$4380, 4820$-$4900, 4920$-$5050, 6500$-$6620, and 6670$-$6770\angs.}, we simultaneously fit the scaled 2001 spectrum with photometric measurements derived from the 2011 XMM-Newton OM images in the UVM2, UVW1, and $V$ bands, 2MASS images in the $J$, $H$, and $K_s$ bands, and WISE images in the W1, W2, W3, and W4 bands.  Our model consists of three components: (1) a stellar component, represented by a \cite{Bruzual2003MNRAS} stellar population synthesis model consisting of two simple stellar populations (SSPs) of solar metallicity\footnote{ The best-fit result with only one SSP left a strong residual on the blue part of the optical spectrum and the XMM-Newton OM UVM2 and UVW1 points, requiring a composite stellar population to achieve a decent fit. Our tests indicate that adding additional SSPs does not improve the fit significantly. }, with allowance for internal extinction; (2) an AGN component, represented by a power law, for scattered emission from the accretion disk, which can be substantial for type~2 AGNs \citep{Bessiere2017MNRAS,Zhao2019ApJ}; and (3) a torus component, using the clumpy CAT3D torus model of \cite{Honig2017ApJ}, but without including the possible effect of a polar wind, which would be difficult to constrain with the currently limited data.  The best-fit parameters are given in Table~\ref{tab:host}.

AGN activity was already present in 1ES\,1927+654 before the outburst. From the pre-outburst 2--10\,keV luminosity of $L_\mathrm{2-10keV} \simeq 2.4\times 10^{42}\rm\,erg\,s^{-1}$ (\citealp{Gallo2013MNRAS}), the empirical correlation between hard X-ray and mid-IR emission \citep{Asmus2015MNRAS} predicts a monochromatic 12 \mum\ luminosity $L_\mathrm{12\, \mu m} = 5.35_{-2.79}^{+5.83}\times 10^{42}\rm\,erg\,s^{-1}$, which is roughly consistent with our photometric measurements in the W3 band ($L_\mathrm{12\, \mu m}^\mathrm{obs} = 2.50\pm1.30 \times 10^{42}\rm\,erg\,s^{-1}$), especially considering the non-simultaneity of the X-ray and mid-IR observations.

The host galaxy stellar mass helps to constrain the BH mass (Section~\ref{sec:BHmass}). From the 2-D image analysis (Table~\ref{tab:photometry}), the integrated $K_s$-band luminosity of 1ES\,1927+654 is $L_{K_s} = 1.10_{-0.11}^{+0.12} \times 10^{10}\, L_{\odot,{K_s}}$. Considering the rest-frame color $(B-V)_0 = 0.42$ mag calculated from the 2001 optical spectrum, we infer a mass-to-light ratio $M/L_K = 0.32$ (scatter 0.05 dex) following the relation reported in \citet{Kormendy2013ARAA}, or $M_\ast = 3.56_{-0.35}^{+0.38} \times 10^{9}\, M_\odot$, a factor of $\sim 2$ larger than the total stellar mass derived from our SED fitting ($M_\ast = 1.27_{-0.15}^{+0.20} \times 10^{9}\, M_\odot$; Table~\ref{tab:host}).  This level of disagreement is not unexpected when comparing current stellar population synthesis models in the optical and near-IR \citep{Baldwin2018MNRAS}, a problem that is also evident in the systematic mismatch of the 2MASS points in our SED fit (Figure\,\ref{fig:hostga}).  The low stellar mass of the host galaxy of 1ES\,1927+654---comparable to that of the Large Magellanic Cloud ($2.7\times 10^{9}\, M_\odot$; van~der~Marel et al. 2002)---suggests that its bulge (Section~\ref{sec:photoOBs}) likely belongs to the pseudo variety (Kormendy \& Kennicutt 2004). According to \cite{Gao2020ApJS}, this holds for all bulges hosted by galaxies with $M_\ast \simeq 10^{8.5}-10^{9.5}\, M_\odot$ in the Carnegie-Irvine Galaxy Survey (\citealp{cgs}), as well as for most galaxies of similarly blue optical colors ($B-V = 0.4$ mag; \citealt{Gao2020ApJS}) and bulge S\'ersic indices ($n\simeq 1.5$; \citealt{Gao2019ApJS}).  We conclude that 1ES\,1927+654 likely hosts a pseudo bulge.

The two SSPs both have best-fit ages less than 1 Gyr, suggesting that 1ES\,1927+654 has undergone a recent starburst. The majority of post-starburst galaxies in the mass range $M_\ast \approx 10^{9.5}-10^{10.5}\,M_\odot$ have experienced a disruptive event such as a gas-rich major merger \citep{Pawlik2018MNRAS}, which may help explain 1ES\,1927+654's somewhat unusually large $B/T$ (Section~\ref{sec:GALFIT}).  It would also reinforce the argument that this changing-look event is associated with a TDE \citep{Trakhtenbrot2019ApJ}, which preferentially occur in post-starburst environments \citep{Arcavi2014ApJ}.

\subsection{After the Outburst}
\label{sec:specmodelafter}

The presence of a featureless blue continuum and broad Balmer lines after the optical outburst indicates that 1ES\,1927+654 changed from a type~2 to a type~1 AGN. Among the 34 optical spectra available, 31 have full wavelength coverage from 4000 to 8000\angs, while the earliest three spectra covered only 4000 to 6700\angs\ and did not include the entire \ha\ profile. The detailed analysis of the spectra in this phase requires a different approach from that used for the spectrum before the outburst (Section~\ref{sec:beforeout}). 
The primary aim is to quantify the evolution of the featureless continuum and broad emission lines, including their luminosities and shapes. We also use this information to estimate the virial BH mass ($M_\mathrm{BH}$; Section~\ref{sec:BHmass}).

\begin{figure}
\centering
\includegraphics[width=0.48\textwidth]{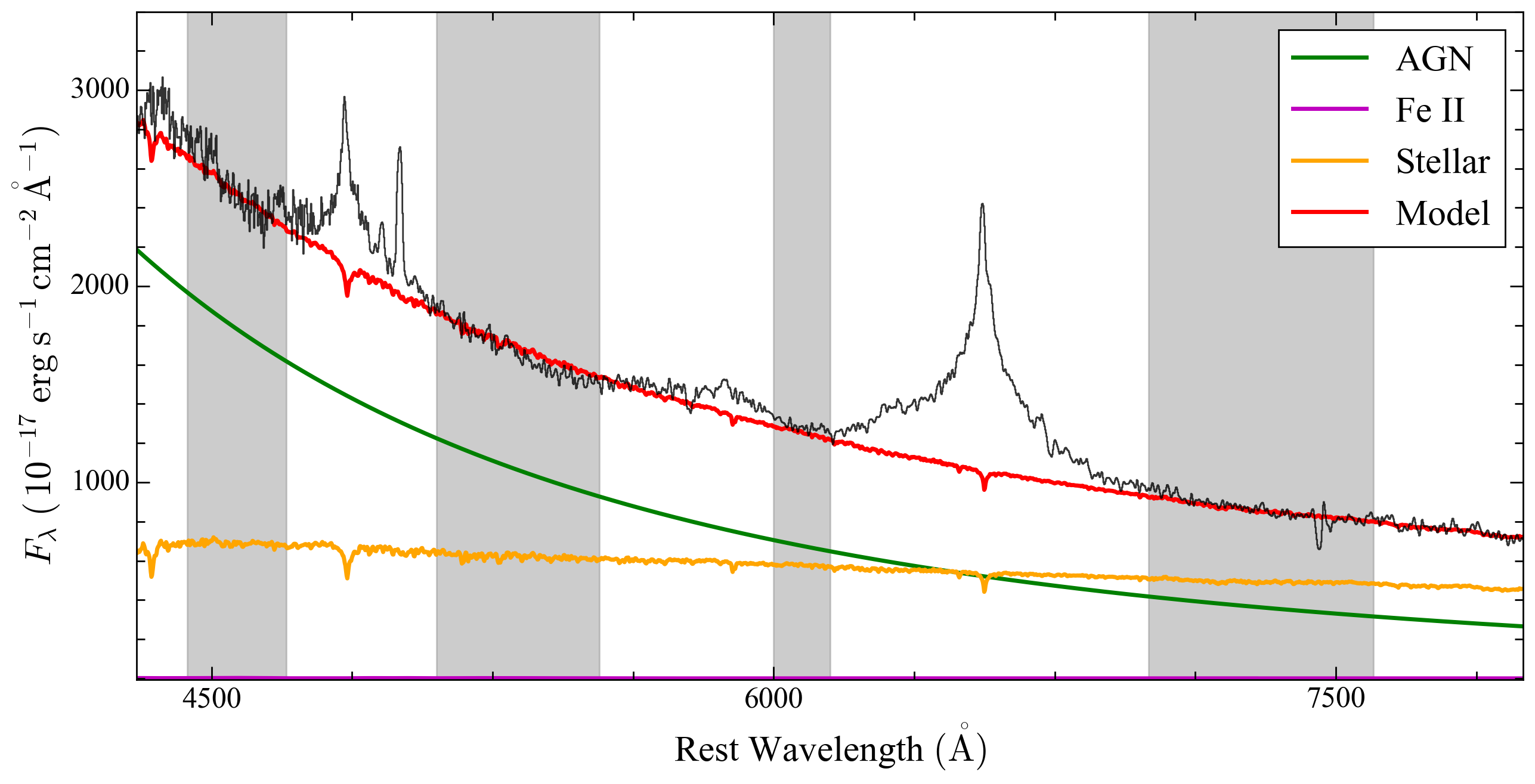}
\caption{Illustration of optical continuum fitting, using as an example the spectrum acquired on 24 April 2018.  The continuum model consists of an AGN  power-law component (green), broad iron emission (purple), and host galaxy starlight (cyan). The overall model (black) reproduces well the spectrum in the four fitting windows marked in gray shaded regions.}
\label{fig:continuummodel}
\end{figure}

The overall continuum must be modeled and subtracted prior to analyzing the emission lines.  After correcting for Galactic extinction and shifting to the rest frame, we construct a model for the continuum comprising (1) a power law for the accretion disk emission, (2) broad, blended iron emission, and (3) starlight from the host galaxy (Figure\,\ref{fig:continuummodel}). For the power-law component, we allow its slope and normalization to vary for each spectrum. For the iron lines, we adopt the empirical template derived from observations of I~Zw~1 \citep{Boroson1992ApJS}, which covers the wavelength range $\sim 3560-7500$\angs, adjusting its normalization and FWHM, but not its radial velocity shift relative to the systemic velocity of the galaxy \citep{Hu2008ApJ,Hu2012ApJ}. Contrary to common practice (e.g., \citealp{Ho2012ApJ}), here we do not consider the Balmer continuum because it contributes negligibly at wavelengths longer than 4000\angs. The host galaxy contribution comes directly from the results of Section~\ref{sec:beforeout}. The emission of 1ES\,1927+654 prior to the outburst was dominated by the host galaxy, which is given by the best-fitting SSP components in Table~\ref{tab:host}. We allow the normalization to change, since different spectra admit different relative fractions of host galaxy light, depending on the aperture size and seeing conditions.  The continuum model, fit over several line-free windows (4435$-$4700, 5100$-$5535, 6000$-$6150, and 7000$-$7600\angs), is illustrated in Figure\,\ref{fig:continuummodel}.

The emission-line fits consider nearly all the standard optical diagnostic narrow forbidden lines, including \oiii~$\lambda\lambda 4959, 5007$, \oi~$\lambda\lambda 6300, 6364$, \nii~$\lambda\lambda 6548, 6584$, and \sii~$\lambda\lambda 6716, 6731$, as well the permitted lines of \heii~$\lambda 4686$, \hei~$\lambda 5877$, \hg, \hb, and \ha, which can have both a broad and a narrow component. Although the first-order global continuum has been removed, the regions surrounding some weak emission lines (e.g., \hg\ or \hei) need to be adjusted using a local power-law continuum to achieve a satisfactory fit. The narrow components of \ha\ or \hb\ are especially troublesome because they are severely blended with their broad counterparts. Following standard practice (e.g., \citealt{Ho1997ApJS,Ho2009ApJS}), we use a nearby, relatively unblended narrow forbidden line as an empirical profile template, namely \oiii\ in the case of \hb\ and, if sufficiently strong, \sii\ in the case of the \ha+\nii\ complex.  We fix the wavelength separation of the doublets to their laboratory values and the relative amplitudes in the case of transitions that originate from the same energy level \citep{Storey2000MNRAS}. As the emission lines can have complex shapes (e.g., \oiii\ often shows an asymmetric blue wing; \citealp{Greene2005ApJ}), we fit them with as many Gaussian components as necessary to achieve clean residuals. In practice two Gaussians usually suffice for the narrow lines, and three for the broad ones.  The narrow and broad components do not need to share the same centroid.
We construct the model using the Python package {\tt lmfit} and implement the fit using the MCMC method through the Python package {\tt emcee}. Finally, the best-fit values and the $1\,\sigma$ uncertainties are calculated from the median and the $16\%$ and $84\%$ values.

To study quantitatively the evolution of the line profiles, we calculate the velocity shift ($\Delta V$) of each component by integrating the first moment of the flux ($\int \lambda f_\lambda d\lambda$) with respect to the median wavelength (MED) of the narrow component, where MED is defined as the location that divides the line flux equally on both sides. We integrated the first moment over 5 times the median absolute deviate, ${\rm MAD} = \int |\lambda - {\rm MED}|f_\lambda d\lambda \; /\int f_\lambda d\lambda$. For the broad components of \ha\ and \hb, we also compute the symmetry parameter

\begin{equation}
S = \frac{\| f_+(x)\|}{\| f_+(x)\| + \| f_-(x)\|},
\label{equ:symmetry}
\end{equation}

\noindent
where $x$ is the velocity centered on the integrated first moment of the profile, while $f(x)$ is the modeled flux density, $f_+(x) = \frac{1}{2}(f(x)+f(-x))$ is the symmetric part of the profile, and $f_-(x) = \frac{1}{2}(f(x)-f(-x))$ is the antisymmetric part. The norm of the profile is $\| f(x)\| = \int |f(x)| dx$.  The basic properties of the emission lines are summarized in Table~\ref{tab:emissionline}.

\startlongtable
\begin{longrotatetable}
\begin{deluxetable*}{lcccccccccccccccc}
\tabletypesize{\tiny}
\setlength{\tabcolsep}{1pt}
\tablecaption{Properties of the Emission Lines\label{tab:emissionline}}
\tablehead{
      \colhead{Date}             &
      \colhead{$\log L_{5100}$} &
      \colhead{$\log L_\mathrm{[O\,III]}$}&
      \colhead{$\log L_\mathrm{H\beta}^n$} &
      \colhead{$\log L_\mathrm{H\beta}^b$} &
      \colhead{FWHM$_\mathrm{H\beta}^b$}&
      \colhead{$\Delta V_\mathrm{H\beta}^b$} &
      \colhead{$S_\mathrm{H\beta}^b$}&
      \colhead{$\log L_\mathrm{H\alpha}^n$} &
      \colhead{$\log L_\mathrm{H\alpha}^b$} &
      \colhead{FWHM$_\mathrm{H\alpha}^b$}&
      \colhead{$\Delta V_\mathrm{H\alpha}^b$} &
      \colhead{$S_\mathrm{H\alpha}^b$}&
      \colhead{$\log L_\mathrm{He\,II}^n$} &
      \colhead{$\log L_\mathrm{He\,II}^b$} &
      \colhead{$\log L_\mathrm{He\,I}^n$} &
      \colhead{$\log L_\mathrm{He\,I}^b$}     \\
       &
\colhead{($\rm erg\,s^{-1}$})&
\colhead{($\rm erg\,s^{-1}$})&
\colhead{($\rm erg\,s^{-1}$})&
\colhead{($\rm erg\,s^{-1}$})&
\colhead{($\rm km\,s^{-1}$}) &
\colhead{($\rm km\,s^{-1}$}) &
 &
\colhead{($\rm erg\,s^{-1}$})&
\colhead{($\rm erg\,s^{-1}$})&
\colhead{($\rm km\,s^{-1}$} )&
\colhead{($\rm km\,s^{-1}$} )&
&
\colhead{($\rm erg\,s^{-1}$} )&
\colhead{($\rm erg\,s^{-1}$} )&
\colhead{($\rm erg\,s^{-1}$} )&
\colhead{($\rm erg\,s^{-1}$} ) \\
\colhead{(1)} &
\colhead{(2)} &
\colhead{(3)} &
\colhead{(4)} &
\colhead{(5)} &
\colhead{(6)} &
\colhead{(7)} &
\colhead{(8)} &
\colhead{(9)} &
\colhead{(10)} &
\colhead{(11)} & 
\colhead{(12)} &
\colhead{(13)} &
\colhead{(14)} &
\colhead{(15)} &
\colhead{(16)} &
\colhead{(17)} 
}
\startdata
$06-03-2018$ & $42.65^{+0.01}_{-0.03}$ & $39.72^{+0.04}_{-0.04}$ & $39.03^{+0.06}_{-0.08}$ & $39.94^{+0.08}_{-0.09}$ & $8978^{+520}_{-493}$    & $-1993^{+87}_{-57}$     & $0.97^{+0.01}_{-0.02}$ & $39.44^{+0.02}_{-0.02}$ & $40.20^{+0.09}_{-0.10}$ & $8996^{+566}_{-484}$    & $-1993^{+87}_{-57}$     & $0.97^{+0.01}_{-0.02}$ & $\leq 39.22$            & $\leq 40.22$            & $\leq 39.09$            & $\leq 39.87$            \\
$08-03-2018$ & $42.17^{+0.01}_{-0.03}$ & $39.05^{+0.07}_{-0.08}$ & $38.15^{+0.19}_{-0.27}$ & $40.09^{+0.02}_{-0.03}$ & $8905^{+922}_{-1351}$   & $-1963^{+206}_{-83}$    & $0.96^{+0.01}_{-0.08}$ & $38.80^{+0.04}_{-0.05}$ & $39.65^{+0.10}_{-0.12}$ & $8873^{+1018}_{-1420}$  & $-1963^{+206}_{-83}$    & $0.96^{+0.01}_{-0.08}$ & $\leq 38.88$            & $\leq 39.87$            & $\leq 38.73$            & $\leq 39.52$            \\
$09-03-2018$ & $42.96^{+0.01}_{-0.03}$ & $39.78^{+0.05}_{-0.06}$ & $39.15^{+0.17}_{-0.27}$ & $40.01^{+0.19}_{-0.25}$ & $13558^{+2014}_{-2134}$ & $-1608^{+301}_{-236}$   & $0.97^{+0.02}_{-0.04}$ & $39.58^{+0.05}_{-0.06}$ & $40.29^{+0.26}_{-0.52}$ & $13440^{+2032}_{-2242}$ & $-1608^{+301}_{-236}$   & $0.97^{+0.02}_{-0.04}$ & $\leq 39.88$            & $\leq 40.88$            & $\leq 39.67$            & $\leq 40.46$            \\
$13-03-2018$ & $43.23^{+0.01}_{-0.03}$ & $40.01^{+0.16}_{-0.22}$ & $39.92^{+0.20}_{-0.31}$ & $40.31^{+0.30}_{-0.50}$ & $12175^{+2659}_{-2155}$ & $-1510^{+507}_{-317}$   & $0.96^{+0.02}_{-0.06}$ & $39.94^{+0.17}_{-0.25}$ & $41.05^{+0.16}_{-0.21}$ & $12485^{+2513}_{-2321}$ & $-1510^{+507}_{-317}$   & $0.96^{+0.02}_{-0.06}$ & $\leq 39.78$            & $\leq 40.94$            & $\leq 39.72$            & $\leq 40.61$            \\
$23-03-2018$ & $43.22^{+0.01}_{-0.03}$ & $40.06^{+0.16}_{-0.19}$ & $39.85^{+0.22}_{-0.29}$ & $40.73^{+0.17}_{-0.22}$ & $5733^{+4438}_{-2663}$  & $-2965^{+3389}_{-1871}$ & $0.68^{+0.11}_{-0.11}$ & $40.04^{+0.23}_{-0.37}$ & $41.13^{+0.11}_{-0.32}$ & $9635^{+3828}_{-4291}$  & $-2057^{+1178}_{-1066}$ & $0.84^{+0.05}_{-0.06}$ & $\leq 40.30$            & $\leq 41.31$            & $\leq 40.21$            & $\leq 41.01$            \\
$23-04-2018$ & $43.46^{+0.01}_{-0.03}$ & $40.68^{+0.15}_{-0.17}$ & $40.44^{+0.17}_{-0.24}$ & $41.43^{+0.17}_{-0.26}$ & $14787^{+1342}_{-1564}$ & $-1484^{+323}_{-283}$   & $0.97^{+0.02}_{-0.04}$ & $40.64^{+0.12}_{-0.13}$ & $41.61^{+0.15}_{-0.31}$ & $14860^{+1261}_{-1736}$ & $-1484^{+323}_{-283}$   & $0.97^{+0.02}_{-0.04}$ & $\leq 40.59$            & $\leq 41.60$            & $39.77^{+0.29}_{-0.52}$ & $40.54^{+0.34}_{-0.46}$ \\
$24-04-2018$ & $43.41^{+0.01}_{-0.03}$ & $40.38^{+0.04}_{-0.05}$ & $39.89^{+0.03}_{-0.03}$ & $41.17^{+0.03}_{-0.03}$ & $9753^{+628}_{-493}$    & $-1330^{+219}_{-195}$   & $0.88^{+0.01}_{-0.01}$ & $40.41^{+0.02}_{-0.05}$ & $41.48^{+0.02}_{-0.03}$ & $15769^{+516}_{-1065}$  & $-1871^{+67}_{-75}$     & $0.96^{+0.00}_{-0.00}$ & $\leq 39.32$            & $\leq 40.34$            & $39.29^{+0.08}_{-0.10}$ & $40.39^{+0.05}_{-0.06}$ \\
$07-05-2018$ & $44.06^{+0.01}_{-0.03}$ & $41.20^{+0.13}_{-0.16}$ & $40.98^{+0.13}_{-0.22}$ & $41.89^{+0.24}_{-0.50}$ & $6590^{+2461}_{-2025}$  & $-150^{+920}_{-869}$    & $0.82^{+0.07}_{-0.06}$ & $41.19^{+0.10}_{-0.16}$ & $42.32^{+0.06}_{-0.10}$ & $13234^{+2045}_{-2566}$ & $-1354^{+276}_{-248}$   & $0.93^{+0.04}_{-0.04}$ & $\leq 41.06$            & $\leq 42.07$            & $40.31^{+0.27}_{-0.49}$ & $41.20^{+0.27}_{-0.41}$ \\
$14-05-2018$ & $44.12^{+0.01}_{-0.03}$ & $41.23^{+0.13}_{-0.14}$ & $40.90^{+0.21}_{-0.35}$ & $41.99^{+0.20}_{-0.34}$ & $5468^{+1699}_{-1401}$  & $-32^{+564}_{-573}$     & $0.87^{+0.04}_{-0.05}$ & $41.29^{+0.10}_{-0.11}$ & $42.36^{+0.10}_{-0.11}$ & $13650^{+1952}_{-2260}$ & $-1158^{+254}_{-234}$   & $0.94^{+0.03}_{-0.03}$ & $\leq 40.98$            & $\leq 41.99$            & $40.41^{+0.22}_{-0.40}$ & $41.26^{+0.23}_{-0.31}$ \\
$28-05-2018$ & $43.86^{+0.01}_{-0.03}$ & $41.06^{+0.09}_{-0.12}$ & $40.91^{+0.14}_{-0.17}$ & $41.71^{+0.22}_{-0.44}$ & $6233^{+2449}_{-1596}$  & $174^{+729}_{-761}$     & $0.82^{+0.06}_{-0.05}$ & $41.10^{+0.09}_{-0.11}$ & $42.21^{+0.06}_{-0.07}$ & $10471^{+1768}_{-1567}$ & $-728^{+219}_{-259}$    & $0.89^{+0.03}_{-0.02}$ & $\leq 40.78$            & $\leq 41.79$            & $40.19^{+0.23}_{-0.39}$ & $41.06^{+0.26}_{-0.33}$ \\
$03-06-2018$ & $44.10^{+0.01}_{-0.03}$ & $41.25^{+0.15}_{-0.16}$ & $41.11^{+0.18}_{-0.27}$ & $41.89^{+0.23}_{-0.45}$ & $6510^{+2011}_{-1479}$  & $386^{+712}_{-563}$     & $0.86^{+0.06}_{-0.07}$ & $41.35^{+0.12}_{-0.17}$ & $42.33^{+0.11}_{-0.14}$ & $10480^{+2297}_{-2077}$ & $-465^{+288}_{-264}$    & $0.93^{+0.03}_{-0.03}$ & $\leq 41.07$            & $\leq 42.07$            & $40.45^{+0.24}_{-0.44}$ & $41.22^{+0.28}_{-0.37}$ \\
$11-06-2018$ & $43.80^{+0.01}_{-0.03}$ & $41.01^{+0.14}_{-0.16}$ & $40.94^{+0.15}_{-0.17}$ & $41.69^{+0.15}_{-0.24}$ & $5499^{+1726}_{-1134}$  & $482^{+651}_{-616}$     & $0.88^{+0.05}_{-0.05}$ & $41.23^{+0.07}_{-0.08}$ & $42.03^{+0.10}_{-0.12}$ & $14705^{+1636}_{-3052}$ & $-263^{+236}_{-235}$    & $0.95^{+0.03}_{-0.03}$ & $\leq 40.65$            & $\leq 41.66$            & $40.25^{+0.20}_{-0.32}$ & $40.78^{+0.28}_{-0.37}$ \\
$13-06-2018$ & $43.09^{+0.01}_{-0.03}$ & $40.06^{+0.16}_{-0.25}$ & $39.95^{+0.21}_{-0.24}$ & $40.90^{+0.03}_{-0.04}$ & $3760^{+828}_{-591}$    & $224^{+347}_{-429}$     & $0.91^{+0.02}_{-0.02}$ & $40.33^{+0.20}_{-0.18}$ & $41.34^{+0.11}_{-0.18}$ & $8466^{+310}_{-264}$    & $-81^{+77}_{-72}$       & $0.89^{+0.01}_{-0.01}$ & $\leq 39.47$            & $\leq 40.50$            & $39.03^{+0.22}_{-0.38}$ & $40.15^{+0.10}_{-0.10}$ \\
$24-06-2018$ & $42.92^{+0.01}_{-0.03}$ & $40.10^{+0.07}_{-0.06}$ & $39.93^{+0.09}_{-0.12}$ & $40.87^{+0.03}_{-0.04}$ & $4537^{+628}_{-567}$    & $1155^{+288}_{-295}$    & $0.89^{+0.03}_{-0.03}$ & $40.31^{+0.03}_{-0.05}$ & $41.22^{+0.05}_{-0.06}$ & $9115^{+337}_{-356}$    & $20^{+80}_{-89}$        & $0.91^{+0.01}_{-0.01}$ & $\leq 39.41$            & $\leq 40.43$            & $39.19^{+0.15}_{-0.23}$ & $40.02^{+0.11}_{-0.12}$ \\
$28-06-2018$ & $42.97^{+0.01}_{-0.03}$ & $40.16^{+0.12}_{-0.11}$ & $40.02^{+0.27}_{-0.29}$ & $40.87^{+0.10}_{-0.12}$ & $5104^{+1814}_{-1037}$  & $957^{+468}_{-490}$     & $0.91^{+0.03}_{-0.03}$ & $40.33^{+0.26}_{-0.23}$ & $41.24^{+0.05}_{-0.05}$ & $8329^{+876}_{-806}$    & $86^{+134}_{-133}$      & $0.97^{+0.01}_{-0.02}$ & $\leq 39.63$            & $\leq 40.64$            & $39.28^{+0.18}_{-0.29}$ & $39.86^{+0.19}_{-0.19}$ \\
$06-07-2018$ & $43.19^{+0.01}_{-0.03}$ & $40.37^{+0.05}_{-0.05}$ & $40.17^{+0.05}_{-0.06}$ & $41.13^{+0.02}_{-0.02}$ & $3958^{+271}_{-234}$    & $892^{+179}_{-161}$     & $0.87^{+0.02}_{-0.02}$ & $40.66^{+0.03}_{-0.04}$ & $41.53^{+0.02}_{-0.03}$ & $8658^{+328}_{-301}$    & $196^{+53}_{-45}$       & $0.94^{+0.01}_{-0.01}$ & $\leq 39.47$            & $\leq 40.48$            & $39.32^{+0.15}_{-0.20}$ & $40.25^{+0.06}_{-0.06}$ \\
$16-07-2018$ & $42.85^{+0.01}_{-0.03}$ & $39.78^{+0.05}_{-0.07}$ & $40.00^{+0.09}_{-0.13}$ & $40.74^{+0.01}_{-0.01}$ & $5043^{+160}_{-148}$    & $1607^{+93}_{-240}$     & $0.85^{+0.02}_{-0.01}$ & $40.34^{+0.07}_{-0.12}$ & $41.23^{+0.02}_{-0.07}$ & $9681^{+164}_{-100}$    & $709^{+18}_{-16}$       & $0.97^{+0.00}_{-0.00}$ & $\leq 38.93$            & $\leq 39.94$            & $39.22^{+0.03}_{-0.03}$ & $40.06^{+0.02}_{-0.02}$ \\
$17-07-2018$ & $43.02^{+0.01}_{-0.03}$ & $40.14^{+0.28}_{-0.19}$ & $39.95^{+0.13}_{-0.14}$ & $40.93^{+0.03}_{-0.03}$ & $4975^{+1572}_{-610}$   & $2393^{+231}_{-1426}$   & $0.73^{+0.04}_{-0.02}$ & $40.38^{+0.12}_{-0.09}$ & $41.33^{+0.04}_{-0.05}$ & $8297^{+1187}_{-561}$   & $441^{+69}_{-56}$       & $0.95^{+0.01}_{-0.01}$ & $\leq 39.30$            & $\leq 40.31$            & $39.30^{+0.10}_{-0.12}$ & $40.07^{+0.08}_{-0.08}$ \\
$27-07-2018$ & $43.65^{+0.01}_{-0.03}$ & $40.79^{+0.12}_{-0.10}$ & $40.57^{+0.11}_{-0.16}$ & $41.57^{+0.05}_{-0.06}$ & $4340^{+863}_{-641}$    & $2601^{+544}_{-565}$    & $0.74^{+0.04}_{-0.03}$ & $40.99^{+0.06}_{-0.07}$ & $41.89^{+0.08}_{-0.09}$ & $9224^{+803}_{-732}$    & $234^{+110}_{-114}$     & $0.95^{+0.01}_{-0.01}$ & $\leq 40.31$            & $\leq 41.31$            & $39.90^{+0.16}_{-0.25}$ & $40.55^{+0.22}_{-0.30}$ \\
$11-08-2018$ & $42.75^{+0.01}_{-0.03}$ & $40.08^{+0.08}_{-0.13}$ & $39.98^{+0.08}_{-0.12}$ & $40.69^{+0.07}_{-0.11}$ & $5240^{+530}_{-507}$    & $1290^{+206}_{-182}$    & $0.93^{+0.04}_{-0.04}$ & $40.41^{+0.07}_{-0.07}$ & $41.15^{+0.15}_{-0.27}$ & $9512^{+598}_{-625}$    & $1252^{+112}_{-117}$    & $0.95^{+0.01}_{-0.01}$ & $\leq 39.54$            & $\leq 40.55$            & $39.19^{+0.16}_{-0.25}$ & $40.22^{+0.12}_{-0.14}$ \\
$12-08-2018$ & $42.90^{+0.01}_{-0.03}$ & $40.13^{+0.17}_{-0.31}$ & $39.89^{+0.23}_{-0.29}$ & $40.74^{+0.13}_{-0.21}$ & $4845^{+1210}_{-1097}$  & $935^{+532}_{-623}$     & $0.89^{+0.05}_{-0.04}$ & $40.51^{+0.13}_{-0.22}$ & $41.16^{+0.12}_{-0.19}$ & $8896^{+1132}_{-1108}$  & $560^{+192}_{-174}$     & $0.93^{+0.02}_{-0.02}$ & $\leq 39.64$            & $\leq 40.65$            & $39.33^{+0.18}_{-0.31}$ & $40.04^{+0.21}_{-0.32}$ \\
$08-09-2018$ & $42.89^{+0.01}_{-0.03}$ & $40.05^{+0.14}_{-0.18}$ & $40.19^{+0.09}_{-0.14}$ & $40.62^{+0.24}_{-0.43}$ & $7375^{+2160}_{-1502}$  & $981^{+278}_{-257}$     & $0.94^{+0.03}_{-0.03}$ & $40.52^{+0.08}_{-0.10}$ & $41.10^{+0.11}_{-0.23}$ & $7439^{+2357}_{-1721}$  & $981^{+278}_{-257}$     & $0.94^{+0.03}_{-0.03}$ & $\leq 39.85$            & $\leq 40.86$            & $39.37^{+0.23}_{-0.38}$ & $39.94^{+0.32}_{-0.43}$ \\
$13-11-2018$ & $43.54^{+0.01}_{-0.03}$ & $40.79^{+0.16}_{-0.16}$ & $40.91^{+0.07}_{-0.13}$ & $41.17^{+0.24}_{-0.48}$ & $10133^{+1777}_{-1878}$ & $1266^{+326}_{-334}$    & $0.95^{+0.03}_{-0.03}$ & $41.27^{+0.07}_{-0.09}$ & $41.76^{+0.06}_{-0.07}$ & $10225^{+1877}_{-2196}$ & $1266^{+326}_{-334}$    & $0.95^{+0.03}_{-0.03}$ & $\leq 40.45$            & $\leq 41.47$            & $40.21^{+0.15}_{-0.23}$ & $40.52^{+0.34}_{-0.49}$ \\
$19-03-2019$ & $42.70^{+0.01}_{-0.03}$ & $40.02^{+0.10}_{-0.10}$ & $39.95^{+0.08}_{-0.08}$ & $40.21^{+0.25}_{-0.49}$ & $10873^{+771}_{-709}$   & $1002^{+233}_{-249}$    & $0.95^{+0.03}_{-0.03}$ & $40.40^{+0.05}_{-0.06}$ & $40.86^{+0.04}_{-0.04}$ & $10914^{+749}_{-730}$   & $1002^{+233}_{-249}$    & $0.95^{+0.03}_{-0.03}$ & $\leq 39.57$            & $\leq 40.58$            & $39.42^{+0.12}_{-0.18}$ & $39.62^{+0.32}_{-0.48}$ \\
$06-04-2019$ & $42.62^{+0.01}_{-0.03}$ & $40.07^{+0.07}_{-0.08}$ & $39.98^{+0.05}_{-0.05}$ & $40.36^{+0.06}_{-0.06}$ & $4142^{+2061}_{-1339}$  & $955^{+99}_{-97}$       & $0.93^{+0.02}_{-0.04}$ & $40.37^{+0.06}_{-0.06}$ & $40.88^{+0.03}_{-0.02}$ & $4192^{+1946}_{-1379}$  & $955^{+99}_{-97}$       & $0.93^{+0.02}_{-0.04}$ & $38.97^{+0.15}_{-0.22}$ & $40.15^{+0.07}_{-0.09}$ & $39.47^{+0.05}_{-0.05}$ & $39.83^{+0.16}_{-0.26}$ \\
$19-05-2019$ & $42.67^{+0.01}_{-0.03}$ & $40.04^{+0.08}_{-0.09}$ & $39.91^{+0.05}_{-0.07}$ & $40.23^{+0.13}_{-0.15}$ & $8996^{+484}_{-1548}$   & $1476^{+174}_{-206}$    & $0.92^{+0.03}_{-0.04}$ & $40.37^{+0.04}_{-0.08}$ & $40.84^{+0.03}_{-0.04}$ & $8996^{+494}_{-1370}$   & $1476^{+174}_{-206}$    & $0.92^{+0.03}_{-0.04}$ & $39.20^{+0.14}_{-0.19}$ & $40.01^{+0.12}_{-0.16}$ & $39.43^{+0.08}_{-0.10}$ & $39.54^{+0.30}_{-0.37}$ \\
$08-06-2019$ & $42.62^{+0.01}_{-0.03}$ & $40.04^{+0.08}_{-0.09}$ & $39.91^{+0.06}_{-0.08}$ & $40.13^{+0.24}_{-0.47}$ & $8193^{+3390}_{-3397}$  & $445^{+1087}_{-1058}$   & $0.86^{+0.08}_{-0.10}$ & $40.39^{+0.04}_{-0.04}$ & $40.83^{+0.06}_{-0.05}$ & $9106^{+540}_{-593}$    & $1400^{+230}_{-225}$    & $0.94^{+0.03}_{-0.03}$ & $39.26^{+0.18}_{-0.27}$ & $39.91^{+0.18}_{-0.38}$ & $39.40^{+0.11}_{-0.15}$ & $39.69^{+0.30}_{-0.45}$ \\
$19-06-2019$ & $42.31^{+0.01}_{-0.03}$ & $39.84^{+0.05}_{-0.05}$ & $39.61^{+0.03}_{-0.04}$ & $39.83^{+0.17}_{-0.25}$ & $7937^{+566}_{-1050}$   & $1424^{+111}_{-123}$    & $0.91^{+0.02}_{-0.03}$ & $40.04^{+0.03}_{-0.04}$ & $40.49^{+0.03}_{-0.03}$ & $7991^{+548}_{-1044}$   & $1424^{+111}_{-123}$    & $0.91^{+0.02}_{-0.03}$ & $39.15^{+0.06}_{-0.07}$ & $39.95^{+0.06}_{-0.07}$ & $38.99^{+0.07}_{-0.08}$ & $39.37^{+0.19}_{-0.23}$ \\
$30-06-2019$ & $42.61^{+0.01}_{-0.03}$ & $40.08^{+0.06}_{-0.08}$ & $39.94^{+0.05}_{-0.04}$ & $40.05^{+0.14}_{-0.15}$ & $8923^{+328}_{-292}$    & $1636^{+110}_{-118}$    & $0.96^{+0.02}_{-0.02}$ & $40.40^{+0.03}_{-0.03}$ & $40.79^{+0.02}_{-0.03}$ & $8969^{+264}_{-337}$    & $1636^{+110}_{-118}$    & $0.96^{+0.02}_{-0.02}$ & $39.43^{+0.07}_{-0.09}$ & $40.15^{+0.08}_{-0.08}$ & $39.40^{+0.06}_{-0.07}$ & $39.88^{+0.16}_{-0.24}$ \\
$03-07-2019$ & $42.67^{+0.01}_{-0.03}$ & $40.02^{+0.08}_{-0.09}$ & $39.96^{+0.07}_{-0.07}$ & $40.06^{+0.25}_{-0.36}$ & $8462^{+598}_{-570}$    & $1601^{+304}_{-256}$    & $0.96^{+0.03}_{-0.05}$ & $40.37^{+0.04}_{-0.05}$ & $40.75^{+0.06}_{-0.06}$ & $8412^{+621}_{-531}$    & $1601^{+304}_{-256}$    & $0.96^{+0.03}_{-0.05}$ & $39.48^{+0.13}_{-0.16}$ & $40.30^{+0.12}_{-0.16}$ & $39.39^{+0.13}_{-0.17}$ & $39.51^{+0.35}_{-0.53}$ \\
$25-07-2019$ & $42.63^{+0.01}_{-0.03}$ & $40.08^{+0.10}_{-0.11}$ & $39.92^{+0.09}_{-0.12}$ & $40.00^{+0.22}_{-0.32}$ & $9215^{+949}_{-1115}$   & $1621^{+337}_{-400}$    & $0.95^{+0.03}_{-0.04}$ & $40.39^{+0.04}_{-0.07}$ & $40.75^{+0.07}_{-0.08}$ & $9234^{+1089}_{-1050}$  & $1621^{+337}_{-400}$    & $0.95^{+0.03}_{-0.04}$ & $39.54^{+0.18}_{-0.23}$ & $40.12^{+0.18}_{-0.28}$ & $39.39^{+0.16}_{-0.27}$ & $39.69^{+0.35}_{-0.53}$ \\
$18-08-2019$ & $42.59^{+0.01}_{-0.03}$ & $40.03^{+0.08}_{-0.08}$ & $39.72^{+0.05}_{-0.07}$ & $40.25^{+0.16}_{-0.19}$ & $8937^{+443}_{-397}$    & $1766^{+181}_{-193}$    & $0.94^{+0.03}_{-0.03}$ & $40.25^{+0.03}_{-0.04}$ & $40.69^{+0.03}_{-0.03}$ & $8950^{+447}_{-412}$    & $1766^{+181}_{-193}$    & $0.94^{+0.03}_{-0.03}$ & $39.49^{+0.08}_{-0.09}$ & $40.00^{+0.12}_{-0.12}$ & $39.20^{+0.10}_{-0.13}$ & $39.63^{+0.25}_{-0.36}$ \\
$19-08-2019$ & $42.56^{+0.01}_{-0.03}$ & $40.05^{+0.04}_{-0.05}$ & $39.80^{+0.03}_{-0.03}$ & $40.00^{+0.05}_{-0.06}$ & $8517^{+406}_{-433}$    & $1731^{+122}_{-147}$    & $0.89^{+0.03}_{-0.02}$ & $40.34^{+0.02}_{-0.03}$ & $40.68^{+0.02}_{-0.02}$ & $8512^{+485}_{-375}$    & $1731^{+122}_{-147}$    & $0.89^{+0.03}_{-0.02}$ & $39.59^{+0.04}_{-0.04}$ & $39.85^{+0.06}_{-0.07}$ & $39.37^{+0.04}_{-0.05}$ & $39.95^{+0.09}_{-0.12}$ \\
$10-09-2019$ & $42.63^{+0.01}_{-0.03}$ & $40.16^{+0.05}_{-0.07}$ & $39.87^{+0.04}_{-0.04}$ & $40.08^{+0.20}_{-0.33}$ & $8366^{+321}_{-365}$    & $2006^{+143}_{-154}$    & $0.96^{+0.02}_{-0.02}$ & $40.39^{+0.03}_{-0.04}$ & $40.63^{+0.06}_{-0.07}$ & $8375^{+319}_{-356}$    & $2006^{+143}_{-154}$    & $0.96^{+0.02}_{-0.02}$ & $39.74^{+0.04}_{-0.05}$ & $39.91^{+0.08}_{-0.10}$ & $39.37^{+0.06}_{-0.08}$ & $39.50^{+0.26}_{-0.36}$ 
 \enddata
\tablecomments{Properties of the emission lines obtained from the MCMC fitting (Section~\ref{sec:specmodelafter}). Col. (1): Date of observations. Col. (2): Monochromatic continuum luminosity at 5100\angs. Cols. (3): Luminosity of \oiii\ $\lambda$5007. Cols. (4)--(5): Luminosity of the narrow and broad components of \hb. Cols. (6)--(8): FWHM, velocity shift, and symmetry of broad \hb. Cols. (9)--(10): Luminosity of the narrow and broad components of \ha. Cols. (11)--(13): FWHM, velocity shift, and symmetry of broad \ha. Cols. (14)--(15): Luminosity of the narrow and broad components of \heii\ $\lambda$4686. Cols. (16)--(17): Luminosity of the narrow and broad components of \hei\ $\lambda$5877. }
\end{deluxetable*}
\end{longrotatetable}

\begin{figure}
\centering
\includegraphics[width=0.48\textwidth]{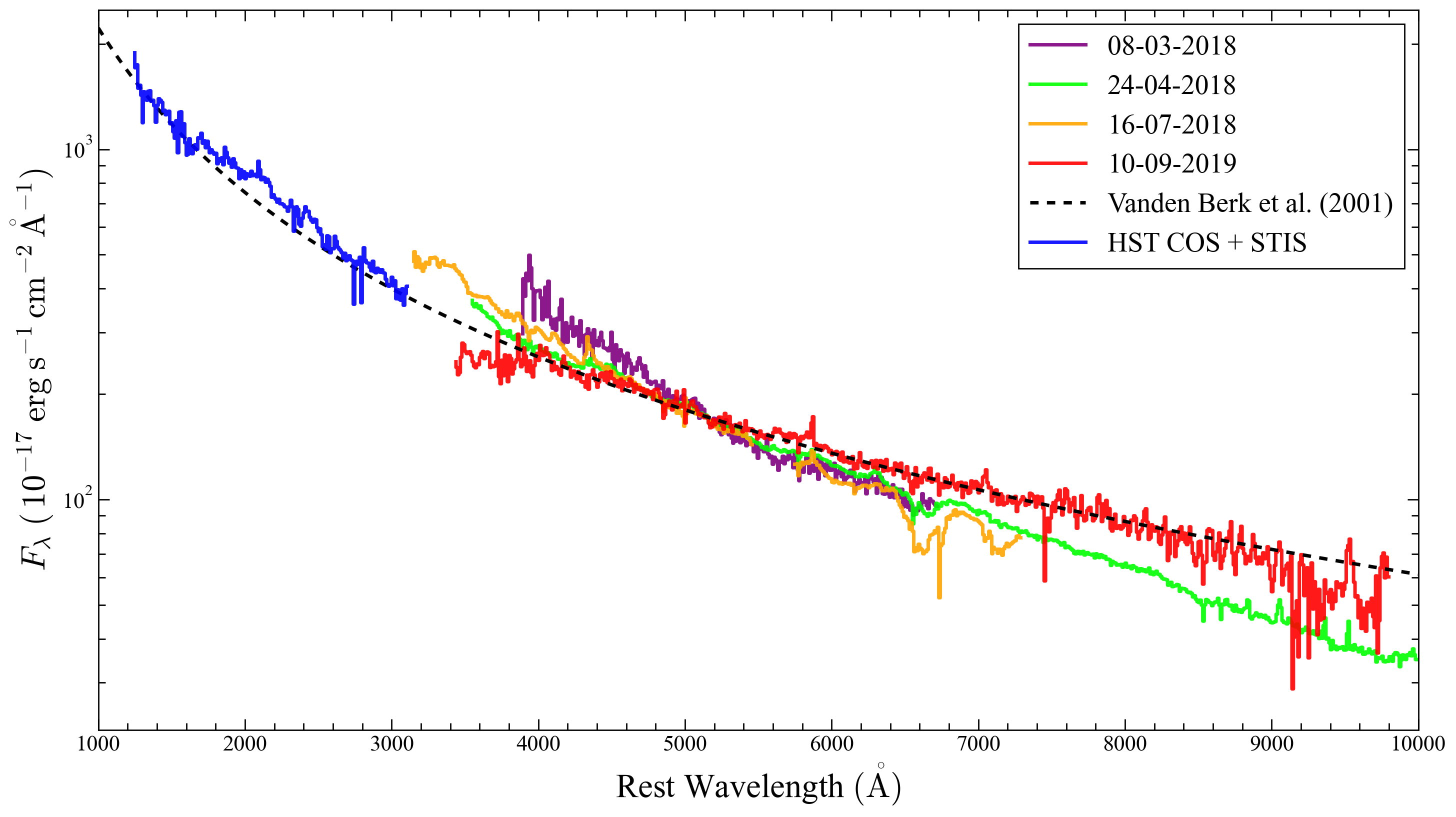}
\caption{Example of continuum fitting of the optical spectra obtained after the outburst (Section~\ref{sec:specmodelafter}). We show spectra taken in four epochs, normalized at 5100\angs\ and after subtracting the emission lines, together with the Hubble Space Telescope (HST) COS and STIS UV spectrum acquired on 28 August 2018 (blue).  The four epochs illustrated here correspond to the colored boxes in  Figure\,\ref{fig:continuum}, as well as the emission-line fits in Figure\,\ref{fig:opdetail}. All the spectra are median rebinned to 10\angs\ per pixel.  The black dashed line represents the power-law fit of the composite spectrum of SDSS quasars \citep{VandenBerk2001AJ}. }
\label{fig:continuumexp}
\end{figure}

\begin{figure}
\centering
\includegraphics[width=0.48\textwidth]{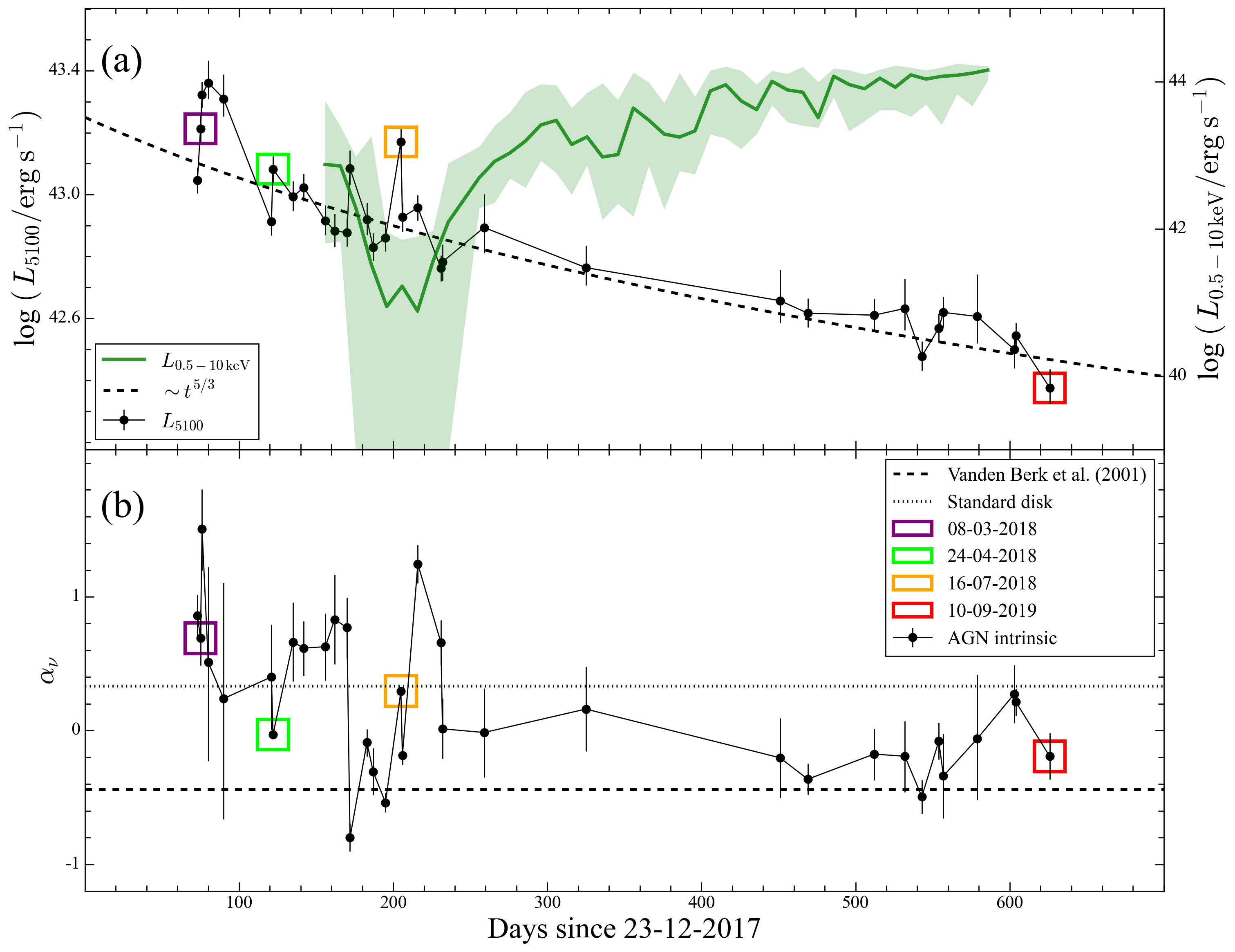}
\caption{Evolution of (a) optical  monochromatic luminosity at 5100\angs\ (black points with error bars; left axis) and X-ray luminosity (green curve; right axis) and (b) optical spectral index after the outburst (Section~\ref{sec:specmodelafter}). The x-axis is in days since the optical outburst (23 December 2017). The optical light curve agrees well with a $\sim t^{-5/3}$ trend (dashed line).  The X-ray light curve, derived from NICER observations \citep{Ricci2020ApJL,Ricci2021ApJS}, is rebinned as the median luminosity every 10 days, and the green shaded region delimits the corresponding minimal and maximal luminosity.  In panel (b), the dashed line denotes the spectral index $\alpha_\nu \simeq -0.44$ derived from the composite spectrum of SDSS quasars \citep{VandenBerk2001AJ}. The dotted line shows the spectral index $\alpha_\nu \simeq 0.33$ predicted by a standard accretion disk \citep{Shakura1973AA}. The colored boxes in both panels highlight the four epochs illustrated in Figures~\ref{fig:continuumexp} and \ref{fig:opdetail}.
 }
\label{fig:continuum}
\end{figure}

\begin{figure}
\centering
\includegraphics[width=0.49\textwidth]{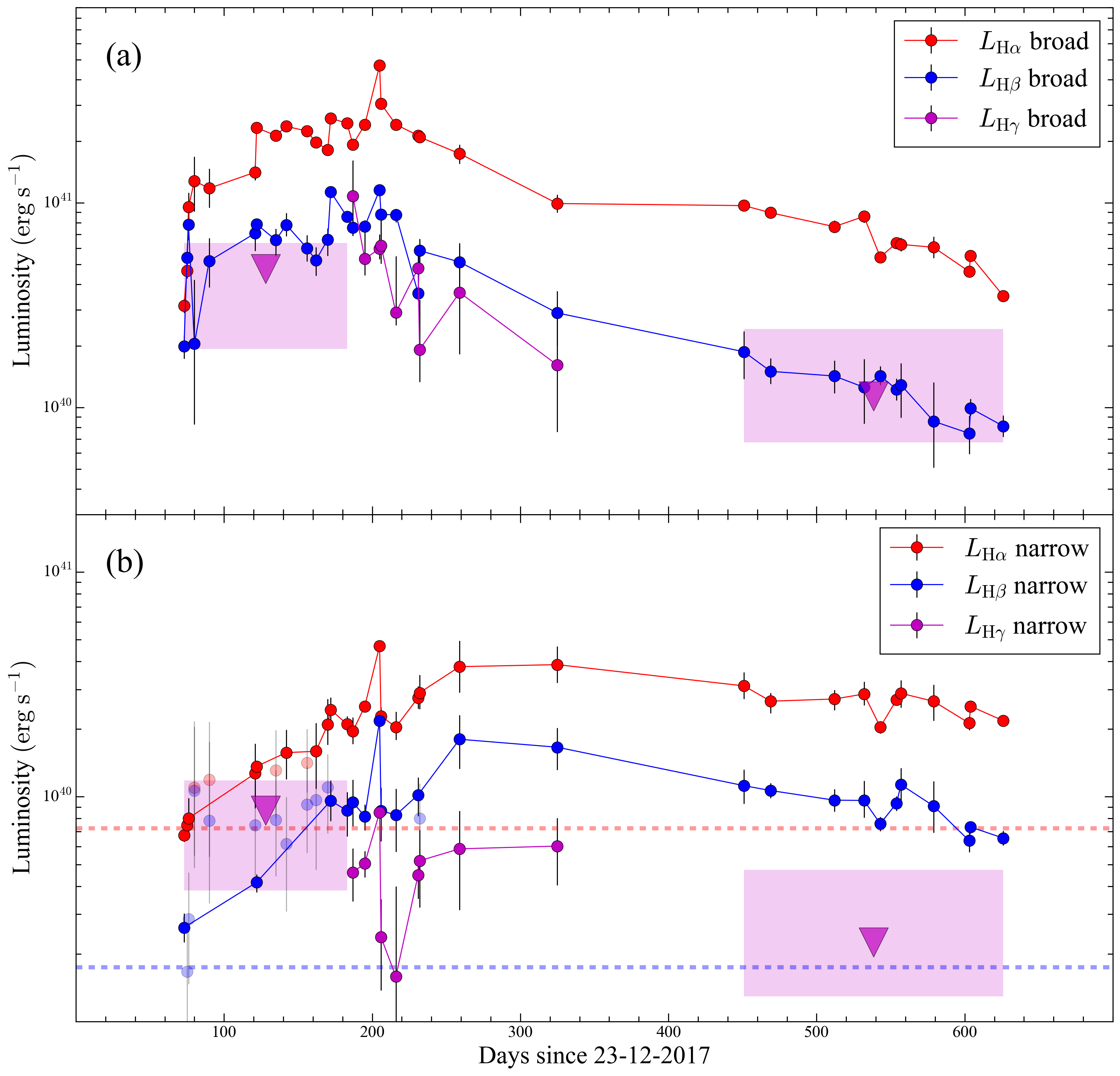}
\caption{Light curves for \ha\ (red), \hb\ (blue), and \hg\ (magenta; detections with ${\rm SNR} > 3$), for (a) the broad component and (b) the narrow component.  The x-axis is in days since the optical outburst (23 December 2017). The non-detections (upper limits) for \hg\  are shown as magenta triangles, where only the median value is plotted and the range is shown as a shaded region. The dashed lines mark the luminosity of narrow \ha\ (red) and \hb\ (blue) luminosity before the optical burst \citep{Boller2003AA}. All the light curves are normalized based on the median \oiii\ luminosity. 
}
\label{fig:hahb}
\end{figure}

\begin{figure*}
\centering
\includegraphics[width=0.96\textwidth]{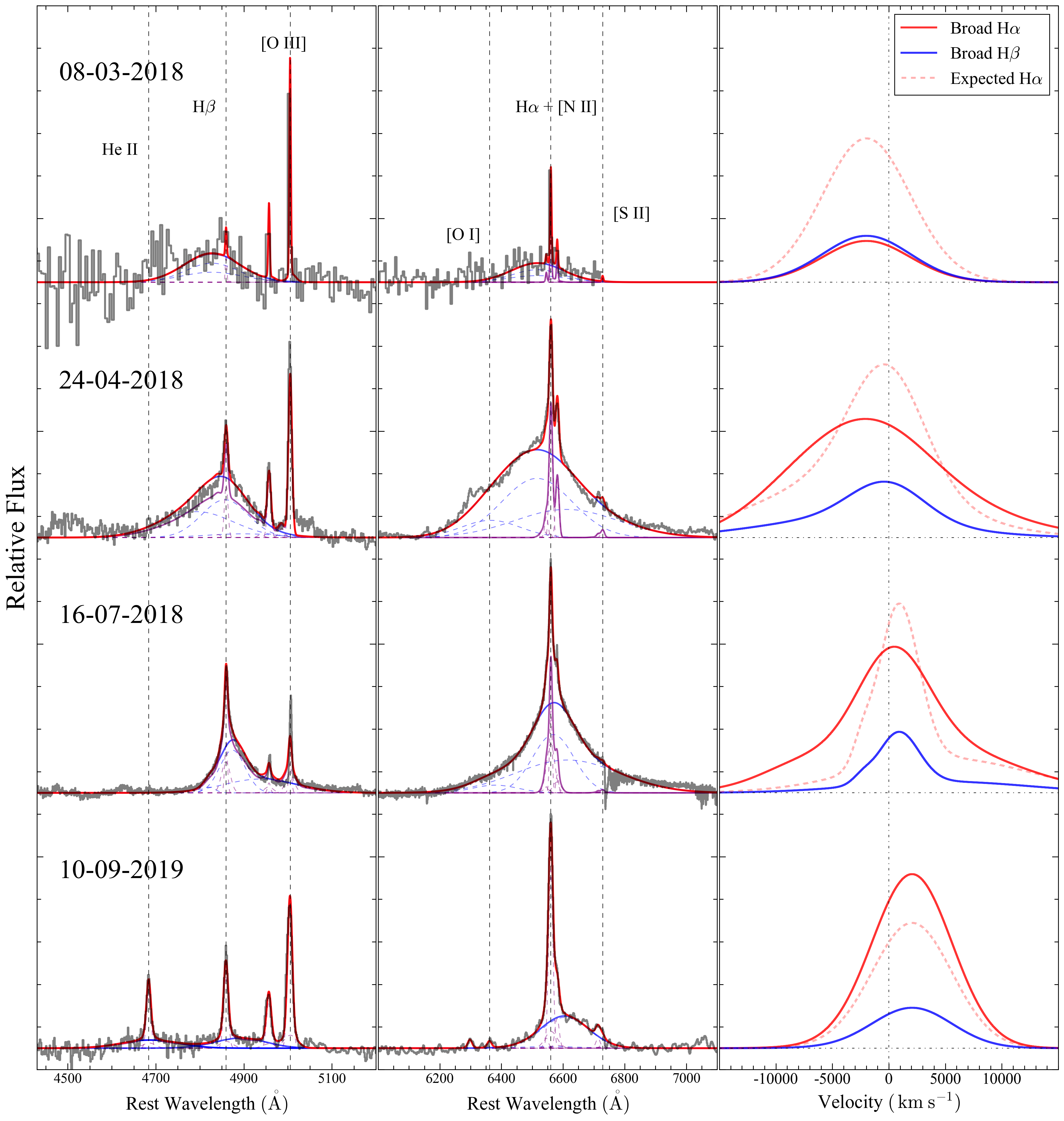}
\caption{Analysis of the emission lines in the \hb\ and \ha\ regions for spectra obtained on 8 March 2018, 24 April 2018, 16 July 2018, and 10 September 2019, where the 8 March 2018 spectrum has been median binned to 5\angs\ per pixel to reduce the noise (for the purposes of the display). The black curve shows the original data, the blue curves represent broad-line components, and the purple curves represent narrow-line components, with individual sub-components further delineated with Gaussians in dashed line. Prominent emission lines are labeled.  The right column shows the best-fit velocity profile of the broad components of \ha\ (solid red curve) and \hb\ (solid blue curve), while the dashed red curve denotes the expected \ha\ line if it had the same profile as \hb\ but with a flux 3.05 times larger, which corresponds to a typical line ratio between \ha\ and \hb\ \citep{Osterbrock2006agnbook}. The black vertical dash-dotted line marks zero velocity shift.}
\label{fig:opdetail}
\end{figure*}

\begin{figure}
\centering
\includegraphics[width=0.49\textwidth]{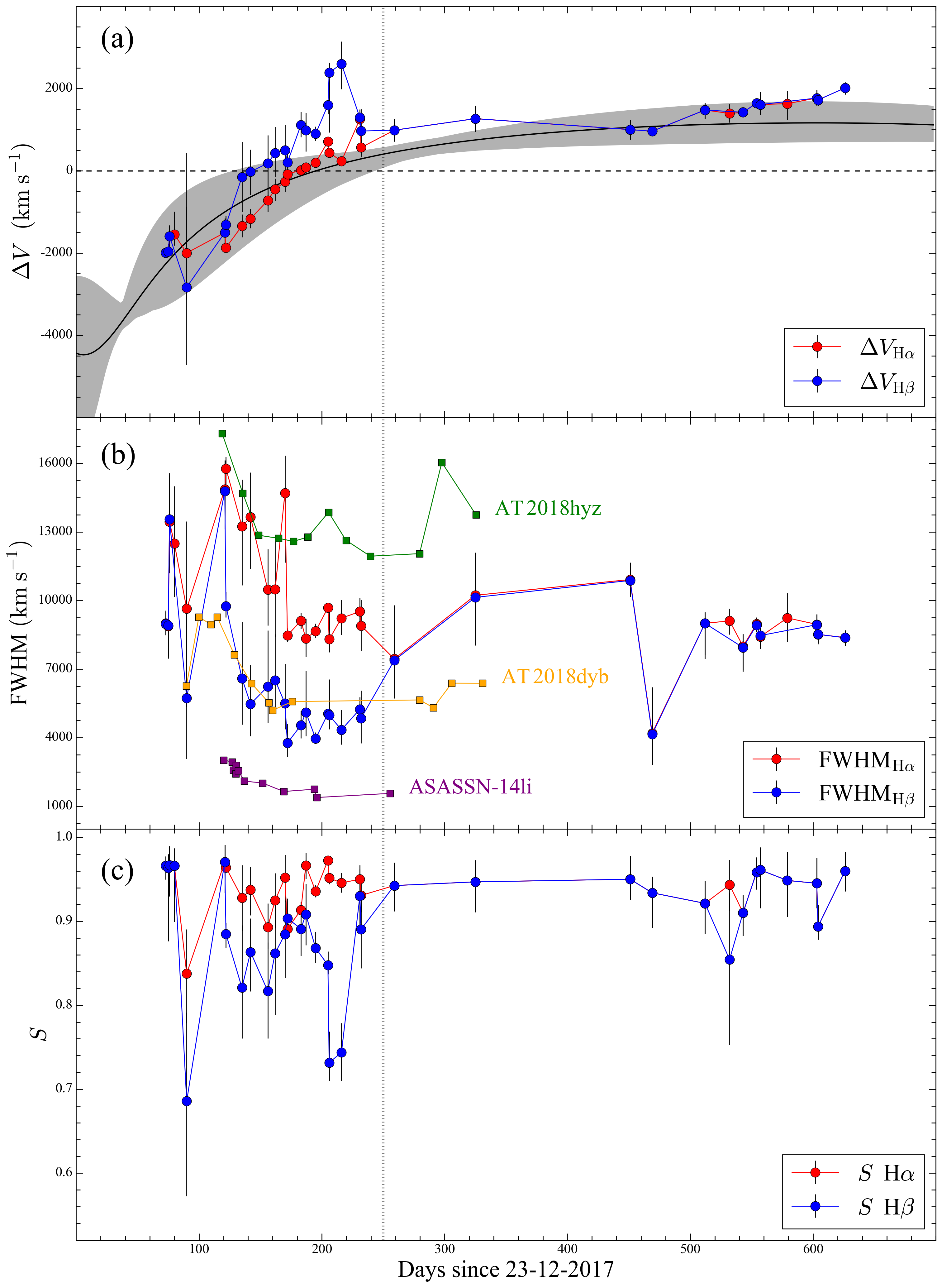}
\caption{Evolution of (a) the velocity shift ($\Delta V$), (b) FWHM, and (c) symmetry parameter ($S$; Equation \ref{equ:symmetry}) of broad \ha\ (red) and \hb\ (blue).  The x-axis is in days since the optical outburst (23 December 2017). The $1\,\sigma$ errors associated with $\Delta V$, FWHM, and $S$ are calculated based on an MCMC method. In panel (a), the overall evolutionary trend was fitted with the Keplerian radial velocity function described by Equation~\ref{equ:kepler}, and the best-fit model and its $1\,\sigma$ uncertainty are plotted as the black curve and gray shaded region, respectively.  In panel (b), the squares show the evolution of FWHM for \ha\ for TDE ASASSN-14li (purple, \citealp{Holoien2016aMNRAS}), AT\,2018dyb (orange, \citealp{Leloudas2019ApJ}), and AT\,2018hyz (green, \citealp{Short2020MNRAS}).  We have shifted the data so that they peak at the same location as 1ES\,1927+654 ($\sim 110$ days). The vertical dotted line marks 250 days after the outburst.}
\label{fig:vshift}
\end{figure}

\section{Spectral Evolution}
\label{sec:specevol}

\subsection{Optical Continuum}

The properties of the intrinsic AGN optical continuum, $f_\nu \propto \nu^{\alpha_\nu}$, can be summarized with the evolution of the monochromatic luminosity at 5100\angs, $L_{5100} \equiv \lambda L_{\lambda}(5100\,{\rm \AA})$, and the spectral index $\alpha_\nu$.  In sharp contrast to the behavior of the X-ray light curve, which decreases from the time of the outburst until $\sim 200$ days and then rebounds \citep{Ricci2020ApJL,Ricci2021ApJS}, $L_{5100}$, consistent with the optical photometric light curve \citep{Trakhtenbrot2019ApJ}, drops monitonically with time approximate as $t^{-5/3}$ (Figure\,\ref{fig:continuum}a).  A disk with a temperature profile $T \propto r^{-p}$ produces a spectrum $f_\nu \propto \nu^{3-2/p}$. For a standard optically thick, geometrically thin accretion disk \citep{Shakura1973AA}, $p=3/4$ and $\alpha_\nu = 1/3$, whereas in a slim disk $p=1/2$ and $\alpha_\nu = -1$ \citep{Kato2008book}.  At the beginning of our spectral series, the value of $\alpha_\nu$ indeed roughly matches that expected for a standard thin disk (Figure\,\ref{fig:continuum}b).  During the nearly 600 days spanned by our observations, the spectral index gradually decreases to a final value of $\alpha_{\nu} \approx -0.3$ to $-0.4$, close to that measured from the Sloan Digital Sky Survey (SDSS) composite spectrum of quasars ($\alpha_\nu = -0.44$; \citealt{VandenBerk2001AJ}).  Interestingly, at $t\approx 200$ days since the optical outburst, $\alpha_\nu$ quickly dropped from the value expected for a standard disk to that observed in SDSS quasars, closely tracking the sharp minimum reached in the X-ray light curve.  Since $\alpha_\nu$ is determined by the temperature profile of the disk, the drop at $t\approx 200$ suggests that the $T \propto r^{-p}$ dependence temporarily became shallower than that of the standard disk solution. The broad-band SED analysis of 1ES\,1927+654 by \citep{Li2022} suggests that efficient cooling suppressed the temperature of the inner disk, resulting in a shallower disk temperature profile and hence $\alpha_\nu$.

\subsection{Behavior of the Broad Balmer Lines}
\label{sec:prochange}

Broad \ha\ and \hb\ are weak but visible in our very first spectrum acquired on 6 March 20186, $\sim 100$ days after the fiducial outburst date of 23 December 2017.  Both lines rose sharply to a maximum that extends from $\sim 120$ to 230 days, with the luminosity of \ha\ reaching $\sim 2 \times 10^{41}\rm\,erg\,s^{-1}$ (Figure\,\ref{fig:hahb}a).  The optical continuum reached its maximum considerably earlier, at approximately 50 days at $\sim 6800$\angs\ ($o$ band of ATLAS), nearly 1--3 months before the first appearance of the broad emission lines \citep{Trakhtenbrot2019ApJ}.  With $L_{5100} \approx  1-3 \times 10^{43}\rm\,erg\,s^{-1}$ during the first 200 days (Figure\,\ref{fig:continuum}a), 1ES\,1927+654 should exhibit a line-to-continuum lag of $\sim 10-16$ days according to the $R_{\rm BLR}-L_{5100}$ relation \citep{Bentz2013ApJ}, or even less if the source is super-Eddington \citep{Du2016bApJ}.  The large disparity between this expected lag and the actually observed time lag suggests that the BLR was formed just after the changing-look event and is probably not yet virialized.  

The most dramatic manifestation of the dynamically evolving nature of the BLR in 1ES\,1927+654 comes from the kinematics of the broad lines.  Figure\,\ref{fig:opdetail} illustrates the evolution of the line profiles, from the first epoch for which we have data (8 March 2018) to the last observation of our campaign (10 September 2019). The profile evolution can be divided into four stages:

\begin{enumerate}
\item{After their emergence around 8 March 2018, the broad components of \ha\ and \hb\ were weak. Both lines have a similar blueshift, and possibly a similar profile, although the low SNR of the lines compel us to keep the two profiles fixed during the fit.}

\item{As the strength of broad \ha\ and \hb\ increased around 24 April 2018, their velocity profiles started to diverge.  We use three Gaussian components to achieve a good fit. \ha\ is notably more blueshifted and blue asymmetric compared to \hb.}

\item{Then, on 2018 July 16, \ha\ and \hb\ were still strong, the lines have significantly different profiles, and, notably, have shifted to positive velocities.}

\item{By 10 September 2019, both \ha\ and \hb\ have become weaker, and \heii~$\lambda 4686$, which has both a broad and narrow component, has emerged. Broad \ha\ and \hb\ are still redshifted, but their line profiles are quite similar. To better constrain the fit, for the epochs in which broad \hb\ was weak, we fixed its profile to that of broad \ha.}
\end{enumerate}

We turn next to the velocity shift ($\Delta V$) of the broad lines as a function of time (Figure\,\ref{fig:vshift}a). The broad \ha\ line is first blueshifted ($\Delta V_{\rm H_\alpha} \approx -2500\, \rm km\, s^{-1}$) and then redshifted ($\Delta V_{\rm H_\alpha} \approx +1500\, \rm km\, s^{-1}$) $\sim 300$ days after the event is thought to have started. Broad \hb\ generally has similar $\Delta V$ as broad \ha, with the largest discrepancy between the two found at $\sim 200$ days, as is already evident from the spectral fits in Figure\,\ref{fig:opdetail}  (cf. second and third rows).  We try to interpret the evolution of $\Delta V$ with a simple Keplerian radial velocity function. Assuming that the evolution is due to the dynamical motion of the emitting clouds,

\begin{equation}
    v_r = \sqrt{\frac{GM_\mathrm{BH}}{a(1-e^2)}}\sin{i} \; [\cos({\phi_{t-t_0}+\psi}) + e\cos{\psi}],
    \label{equ:kepler}
\end{equation}

\noindent
where $v_r$ is the radial velocity of the clouds, $G$ is the gravitational constant, $a$ and $e$ are the semi-major axis and the eccentricity of the orbit, respectively, $\phi_{t-t_0}$ is the true anomaly as a function of time, with $t_0$ the initial time and $t$ the rest-frame time, $i$ is the inclination of the orbit with respect to the observer, and $\psi$ is the longitude of the periapse. We first fix the inclination angle to $i=80^{\circ}$ to minimize parameter degeneracy.\footnote{We tried to fit different values of $i$ and found that $i=80^{\circ}$ best describes the data. Lower $i$ (face-on inclination) would result in values of $a$ too small to match the observed $\Delta V$.}  Fixing $M_\mathrm{BH} = 1.38 \times 10^{6}\, M_\odot$ estimated from the $M_\mathrm{BH} - M_\mathrm{bulge}$ relation (Section~\ref{sec:BHmass}), we use an MCMC method to fit the model, which has free parameters $t_0$, $a$, $e$, and $\psi$, to the observed $\Delta V$ of the broad \ha\ component\footnote{We focus on \ha\ because it has higher SNR compared to \hb\ or \hei, and because the tests described in Appendix~A indicate that the velocity shifts measured from broad \ha, unlike those from broad \hb, are robust against the low spectral resolution of some of our data.} The resulting highly eccentric orbit, with $e=0.59^{+0.22}_{-0.27}$, $a=1.33^{+0.04}_{-0.02}$ light days ($\sim 8500\, R_g$), $t_0 = -16.34^{+6.34}_{-6.30}$, and $\psi = 174.09^{+0.89}_{-0.93}$, covers most of our data points within the $1\,\sigma$ uncertainty (gray shaded region Figure\,\ref{fig:vshift}a). However, the orbital timescale we obtained is $\tau_\mathrm{dyn} \approx 2\pi\sqrt{a/GM_\mathrm{BH}} \approx 1100$ days, about twice as long as the duration of our observations. Therefore, this picture can be confirmed by future spectroscopic observations.
 
The FWHM of \hb\ and \ha\ generally decreases with time (Figure\,\ref{fig:vshift}b). Interestingly, the luminosity of the optical continuum also follows a similar trend (Figure\,\ref{fig:continuum}). This is inconsistent with what is typically observed in type\,1 AGNs, wherein ${\rm FWHM} \propto R_\mathrm{BLR}^{-1/2} \propto L_{5100}^{-0.27}$ (\citealp{Bentz2013ApJ}). A decrease in both the continuum luminosity and the line width is often seen in TDEs, attributed to a decelerating outflow \citep{Yang2013ApJ}.   For comparison, we overplot the evolution of the FWHM of \ha\ for the well-known TDEs ASASSN-14li (purple, \citealp{Holoien2016aMNRAS}), AT\,2018dyb (orange, \citealp{Leloudas2019ApJ}), and AT\,2018hyz (green, \citealp{Short2020MNRAS}), which all have period of broad \ha\ emission detection $> 100$ days after the outburst. To better compare the evolution timescale, we match them to peak at the same location as the \ha\ FWHM of 1ES\,1927+654 ($\sim 110$ days).  The FWHM of \ha\ of 1ES\,1927+654 dropped from $16000\, \rm \, km \, s^{-1}$ to half of its maximum ($\sim 8000\, \rm \, km \, s^{-1}$) in roughly 100 days. This is longer than the $\sim 50$ days it took ASASSN-14li to reach its half maximum, but the value may be underestimated since the observations of ASASSN-14li did not catch the first phase of the \ha\ profile variations.  In spite of their quite different widths, the decreasing timescale of AT\,2018dyb and AT\,2018hyz appears consistent with that of 1ES\,1927+654. More interestingly, all three exhibit an ``echo'' of increasing FWHM after $\sim250$ days, which may be related to the life time of the clouds (see Section~\ref{sec:physpict}). For \hb, it should be noted that while its early and late-time profiles are roughly similar to that of \ha, from $\sim 140$ to 250 days \hb\ is markedly narrower {\it and}\ more asymmetric (Figure\,\ref{fig:vshift}c) compared to \ha. A physical pictre to explain the above phenomenon is presented in Section~\ref{sec:physpict}. Prior to $\sim250$ days, the clouds were actively being formed. BLR clouds with different radial velocity may have different ionization state and density, leading to the production of different fractions of broad \ha\ and \hb.

\subsection{Variability of the Narrow Emission Lines}
\label{sec:narrow_var}

Figure\,\ref{fig:hahb}b reveals a surprising finding: the narrow component of \ha\ and \hb\ vary.  Relative to the historic luminosity of $L_\mathrm{H\alpha} = 7.3\times10^{39}\,\rm erg\,s^{-1}$ and $L_\mathrm{H\beta} = 1.8\times10^{39}\,\rm erg\,s^{-1}$ \citep{Boller2003AA}, during the outburst the narrow components of both \ha\ and \hb\ increased systematically by as much as a factor of 4 after 200 days, and then gradually declined, although by the end of the last epoch monitored the flux had not yet returned to its pre-outburst level. It is notable that the light curves for the narrow lines rise and fall more gradually than those of the broad lines (Figure\,\ref{fig:hahb}a).  The detection of variable narrow-line emission further attests to the complex kinematics and physical conditions of the dynamically evolving debris created in the aftermath of the accretion event that triggered the changing-look event (Section~\ref{sec:physpict}). Interestingly, TDE ASASSN-18pg also showed additional narrow \ha\ emission that emerged and evolved more slowly than the broad compomemt of \ha\ \citep{Holoien2020ApJ} .
However, if narrow \ha\ and \hb\ vary, it is possible that \oiii\ also varies, which would compromise our relative flux calibration strategy that assumes \oiii\ to be constant (Section~\ref{sec:fluxcal}).  We cannot resolve this uncertainty with our present dataset, but we point out that our calibration assumption places a strict lower limit on the actual level of intrinsic flux variations reported in this paper.

\begin{figure}
\centering
\includegraphics[width=0.49\textwidth]{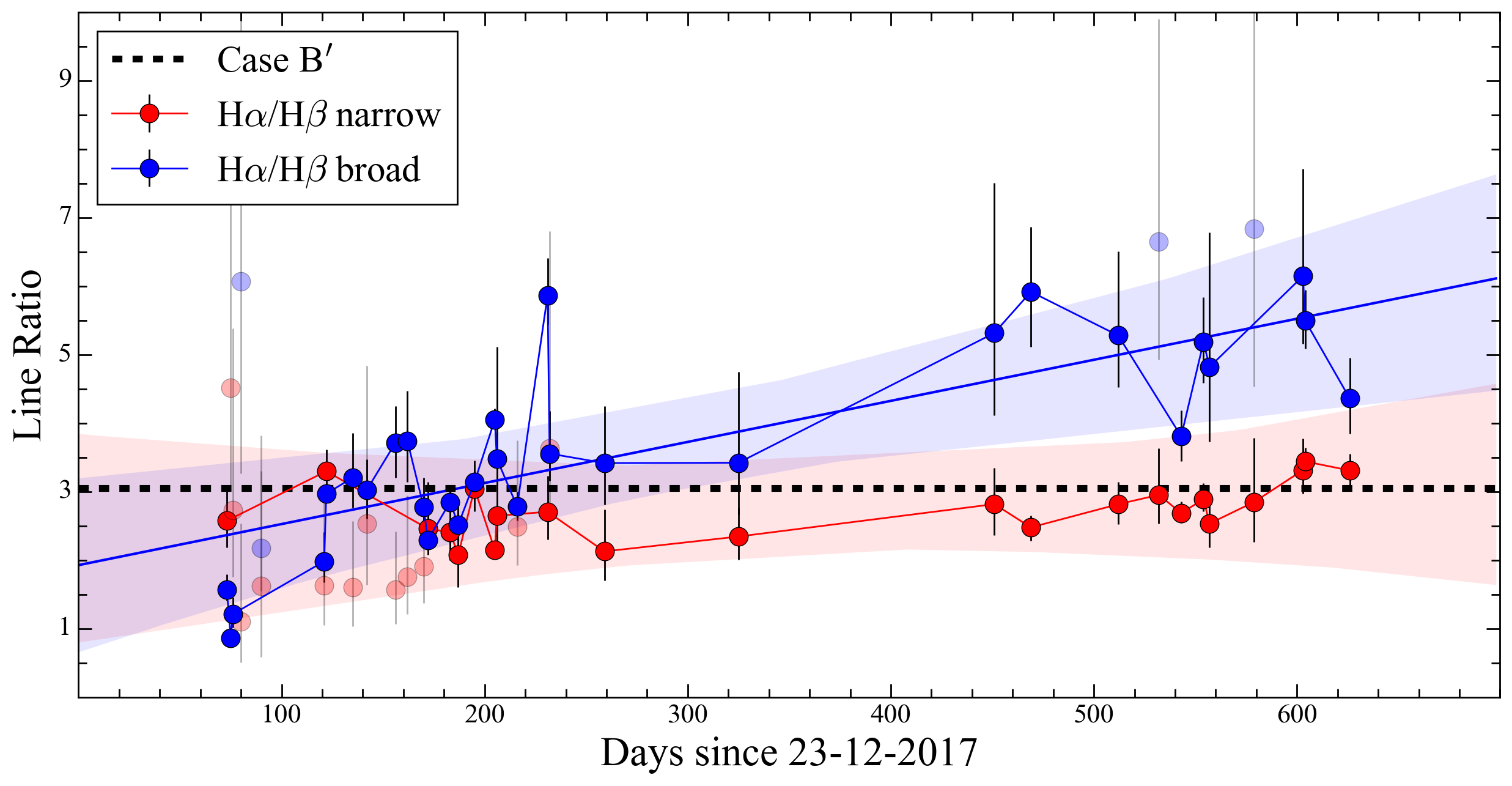}
\caption{Evolution of the Balmer decrement for narrow \ha/\hb\ (red) and broad \ha/\hb\ (blue), as derived from our MCMC analysis, where only measurements with ${\rm SNR}>3$ are shown. A linear fit to the broad lines gives $(\mathrm{H\alpha}/\mathrm{H\beta})_{b} = 3.00\times(t/500\,\mathrm{days})+1.93$ (blue line, shaded region $1\,\sigma$ uncertainty). No significant evolution is observed for narrow \ha/\hb, whose $1\,\sigma$ uncertainty is given by the red shaded region. The black dashed line marks the expected ratio $\mathrm{H\alpha}/\mathrm{H\beta} = 3.05$ for Case B$^\prime$ recombination \citep{Osterbrock2006agnbook}.
}
\label{fig:balmerdecrement}
\end{figure}

\begin{figure}
\centering
\includegraphics[width=0.49\textwidth]{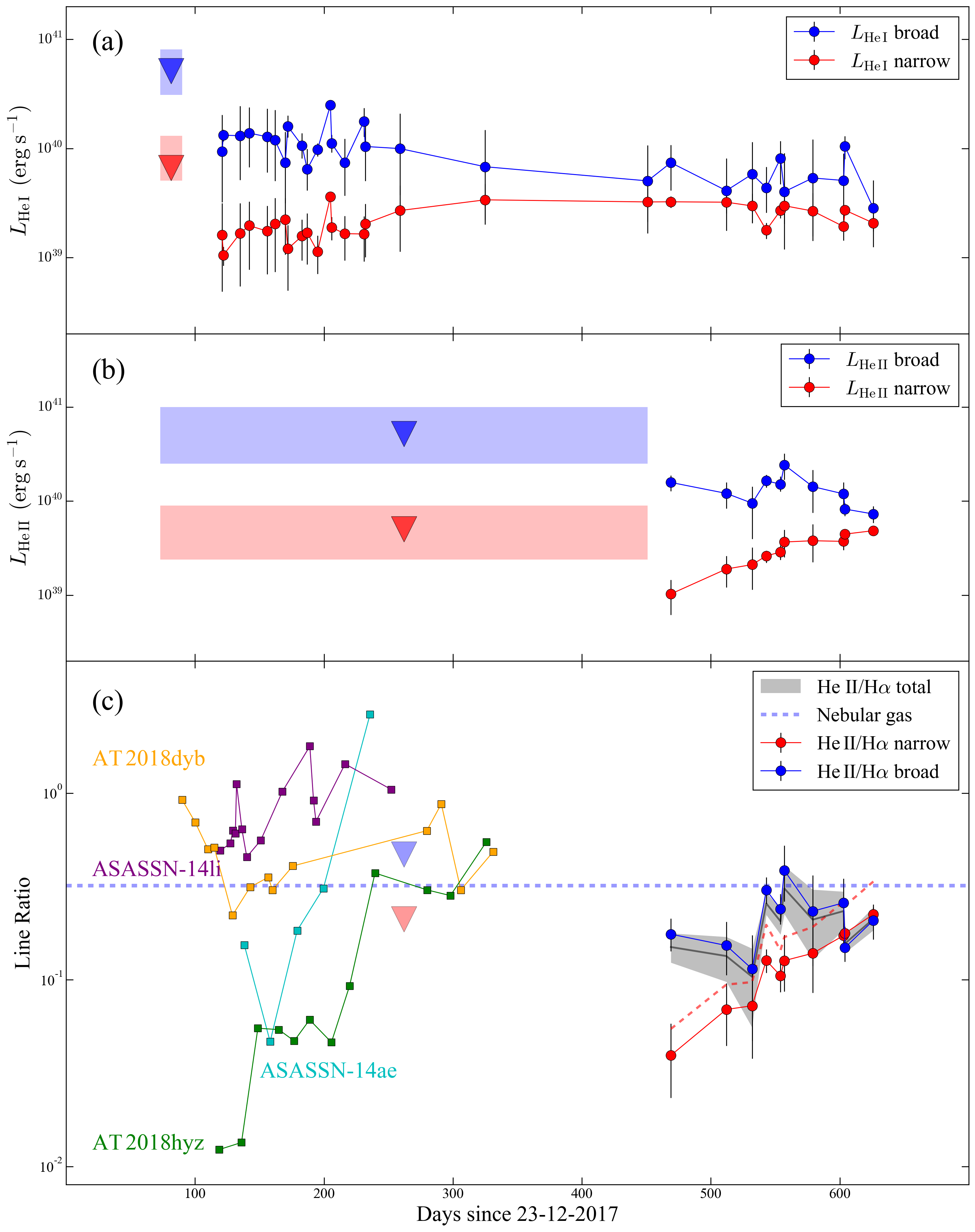}
\caption{Evolution of (a) \hei\ luminosity, (b) \heii\ luminosity, and (c) \heii/\ha\ ratio for the broad (blue points) and narrow (red points) components of the lines.  Upper limits are given when the line is not detected (${\rm SNR} < 2$).  The median value of upper limits is labeled as downward-pointing triangles, with the range shown as a shaded region.  The x-axis is in days since the optical outburst (23 December 2017). In panel (c), the blue dashed line shows \heii/\ha\ $\simeq 0.32$ expected for fully ionized gas with solar helium abundance \citep{Hung2017ApJ}.  The gray curve shows the total \heii/\ha\ ratio obtained by directly adding the flux of the narrow and broad components, while the gray shaded region marks the $1\,\sigma$ (16$\%$ to 84$\%$) confidence range from our MCMC fitting.  The red dashed line gives the narrow \heii/\ha\ ratio after subtracting the original narrow \ha\ flux before the optical outburst \citep{Boller2003AA}.  We also plot the line ratios obtained for a few well-known TDEs (squares): ASASSN-14li (purple) and ASASSN-14ae (cyan) from \citet{Hung2017ApJ}, at 110 days after discovery, and AT\,2018dyb (orange) and AT\,2018hyz (green) from \citet{Charalampopoulos2022AA}, at 110 days from the peak of their light curves. }
\label{fig:heiiuplim}
\end{figure}

\subsection{Balmer Decrement}
\label{sec:balmer_decrement}

Our line decomposition affords us an opportunity to track the time variation of the Balmer decrement (Figure\,\ref{fig:balmerdecrement}).  During the course of the monitoring campaign, the narrow component of \ha\ and \hb\ have a ratio between 2 and 3. The Balmer decrement of the narrow component remains constant at a value consistent with Case B$^\prime$ recombination for low-density photoionized gas (\ha/\hb = 3.05; \citealt{Osterbrock2006agnbook}). The broad component of the Balmer lines behaves differently.  While initially $(\mathrm{H\alpha}/\mathrm{H\beta})_{b}$ is substantially lower than 3.05, perhaps even as low as $\sim 1$, after $\sim 200$ days it increases much more dramatically with time: $(\mathrm{H\alpha}/\mathrm{H\beta})_{b} = (3.00 \pm 0.52)\; \times(t/500\,\mathrm{days})+(1.93\pm0.34)$ (blue points, curve, and shaded region).  By the end of our campaign, $(\mathrm{H\alpha}/\mathrm{H\beta})_{b} \approx 5$, significantly higher than the Case B$^\prime$ value attained by the narrow lines (see the trend also in the third column of Figure\,\ref{fig:opdetail}).  We believe that the systematic time variation of $(\mathrm{H\alpha}/\mathrm{H\beta})_{b}$ and the large values it presents at late times are not a trivial consequence of line-of-sight reddening by dust, which, in any case, appears negligible from the Balmer decrement of the narrow lines.  Instead, the steepening of the Balmer decrement of the broad lines may reflect a decrease of both the evolving density of the clouds and the ionizing photons.  From theoretical considerations, a Balmer decrement as low as $\sim 1$ may arise under conditions of very high densities ($n_\mathrm{H}>10^{12}\rm \,cm^{-3}$; \citealp{Korista2004ApJ}). Under such conditions, the finite number of scatterings between energy levels 3 and 4 of the hydrogen atom and Ly$\beta$ leakage act to suppress the \ha\ intensity, leading to low $(\mathrm{H\alpha}/\mathrm{H\beta})_{b}$ \citep{Netzer1975MNRAS}. However, such a low flux ratio disappears after 150 days for all velocities (third column of Figure\,\ref{fig:opdetail}), suggesting that the clouds newly formed during this phase were not as dense as those in earlier epochs.  At the other extreme, very high values of the Balmer decrement for the broad lines ($\sim 5$) can be attained at late times after the ionizing photon density drops sufficently low ($n_\gamma \approx 5\times 10^{7}\rm \,cm^{-3}$; \citealp{Korista2004ApJ}), resulting in larger Ly$\alpha$ optical depth and hence a steepening of the Balmer decrement \citep{Netzer1975MNRAS,Rees1989ApJ}.

\subsection{The Emergence of the Helium Lines}
\label{sec:heiilines}

As with the Balmer lines, \hei~$\lambda5877$ and \heii~$\lambda4686$ also exhibit both a narrow and a broad component, although they surfaced at different times.  The earliest detection of \hei\ occured at $t \approx 120$ days (Figure\,\ref{fig:heiiuplim}a), but \heii\ did not appear until $t \approx 460$ days (Figure\,\ref{fig:heiiuplim}b). Intriguingly, \hg\ also emerged relatively late ($t\approx 180-320$ days), but unlike the helium lines, it did not persist until our last epoch of observation (Figure\,\ref{fig:hahb}).  It is interesting to note that both the narrow and broad components of these three lines appeared roughly at the same time. By analogy with the emergence of the two kinematically distinct components of \ha\ and \hb\ (Figure\,\ref{fig:hahb}), we surmise that these narrow emission lines were associated with the second kinematic component of the BLR clouds, whose velocity field was dominated by the outflow. Shortly after \hei\ was detected, both the velocity shift and FWHM of its broad component followed those of \hb. We did not detect any velocity shift for \hei\ or \heii\ at late times.  For all the emission lines, the narrow component was only $\sim 10\%$ as bright as the broad emission, which suggests that the outflow clouds comprise only a small fraction of all the clouds.   

The two kinematic components of \hei\ and \heii\ follow a notably different evolutionary trend: whereas the narrow component rises in strength over time, leveling off to a broad maximum and then possibly turning over in the case of \hei, the broad component monotonically decreases with time. \heii\ is commonly found in TDEs (e.g., \citealp{Gezari2012Nature,Hung2017ApJ}), many early in their evolution (e.g. ASASSN-14li and AT\,2018dyb). However, in none of them does \heii\ emerge as late as in 1ES\,1927+654 \citep{Charalampopoulos2022AA}.  It is also interesting to notice that the \heii/\ha\ flux ratio behaves very differently among different objects (Figure\,\ref{fig:heiiuplim}c). In general, the ratio is determined first by the temperature of the medium (recombination efficiency) and second by the abundance ratio between He$^{+2}$ and H$^+$. The median value of \heii/\ha\ is $\sim$0.2 (both narrow and broad components combined; Figure\,\ref{fig:heiiuplim}c), somewhat lower than expected for typical fully ionized nebular gas with solar helium abundance (Y$_\odot=0.2485$; \citealp{Serenelli2010ApJ}), for which \heii/\ha\ $\simeq 0.32$ (\citealp{Hung2017ApJ}). The lower value of narrow \heii/\ha\ likely can be attributed to the contamination of \ha\ emission by the narrow-line region clouds already present prior to the optical outburst. After subtracting the original \ha\ flux, the ``post-outburst" \heii/\ha\ increased significantly (red dashed line; Figure\,\ref{fig:heiiuplim}c) and finally reached the value for fully ionized He (blue dashed line).  This suggests that the photonionized outflowing clouds also decreased in density, in the same manner as the Keplerian bound component, thereby elevating the abundance ratio between He$^{+2}$ and H$^+$ at late times (see also the discussion in Section \ref{sec:physpict}).

\begin{figure}
\centering
\includegraphics[width=0.49\textwidth]{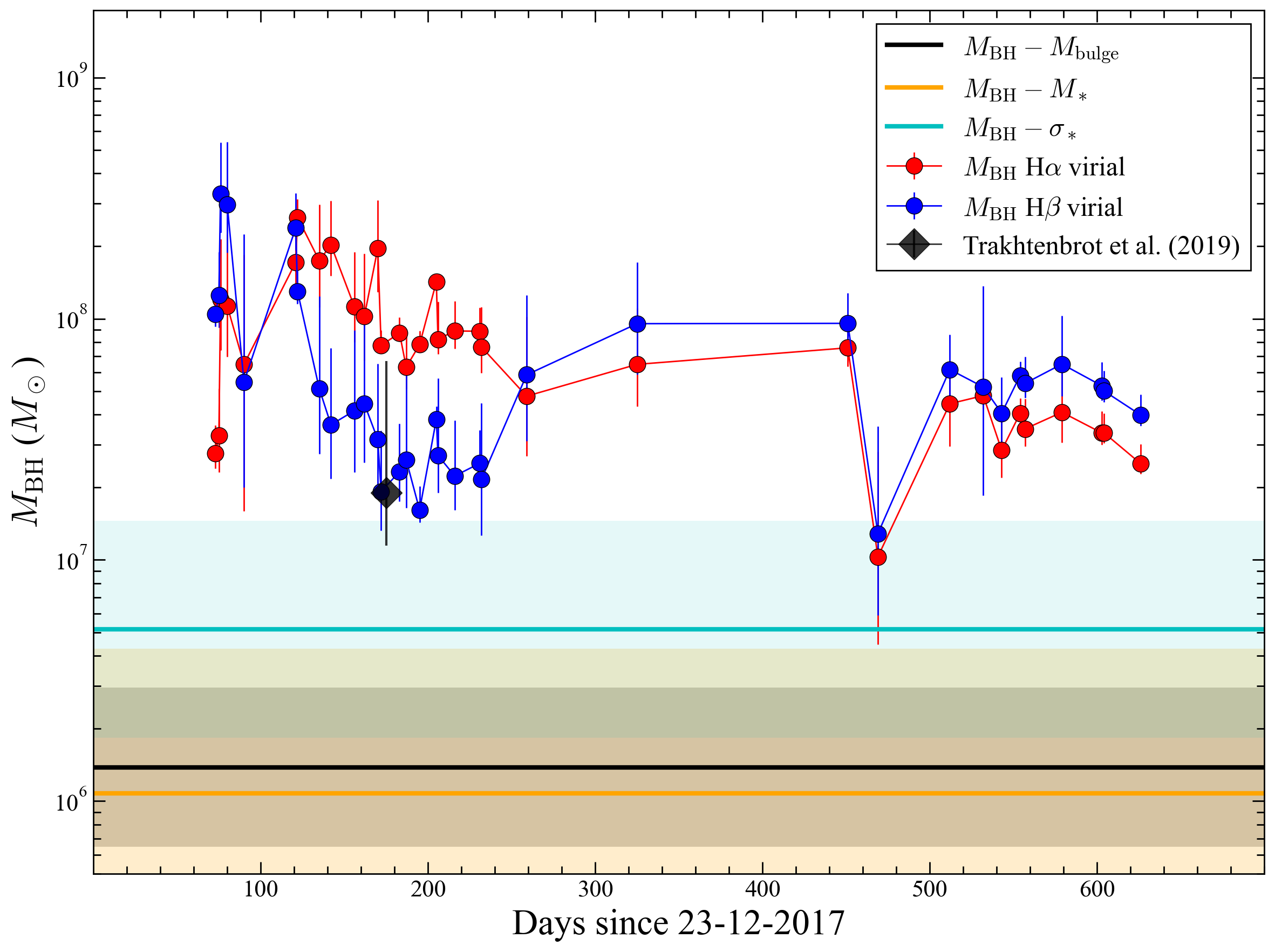}
\caption{Black hole mass ($M_\mathrm{BH}$) estimates for 1ES\,1927+654. The x-axis is in days since the optical outburst (23 December 2017). The red points show the virial $M_\mathrm{BH}$ estimated from broad \ha\ \citep{Greene2005ApJ}, while the blue points correspond to values based on broad \hb\ \citep{Ho2015ApJ}, assuming a virial factor appropriate for pseudo bulges (Section~\ref{sec:beforeout}). The large black point represents $M_\mathrm{BH}$ from the \hb\ profile fit of \citet{Trakhtenbrot2019ApJ}, which is consistent with our results. The three colored horizontal lines and their corresponding shaded regions denote the mass and $1\,\sigma$ confidence interval estimate from different scaling relations: $M_\mathrm{BH} - M_\mathrm{bulge}$ correlation (\citealp{Kormendy2013ARAA}; $M_\mathrm{BH} = 1.38_{-0.73}^{+1.57} \times 10^{6}\, M_\odot$; black), $M_\mathrm{BH} - M_\ast$ correlation (\citealp{Greene2020ARAA}; $M_\mathrm{BH} = 1.08_{-0.81}^{+3.21} \times 10^{6}\, M_\odot$; orange), and $M_\mathrm{BH}-\sigma_\ast$ correlation (Saglia et al. 2016; $M_\mathrm{BH} = 5.1_{-3.2}^{+9.3} \times 10^{6}\, M_\odot$; cyan). 
}
\label{fig:virialMbh}
\end{figure}

\subsection{Black Hole Mass Estimates}
\label{sec:BHmass}

We have at our disposal several methods to estimate the mass of the BH in 1ES\,1927+654.  The availability of multiple epochs of optical spectroscopy in principle may allow us to perform reverberation mapping \citep{Blandford1982ApJ} to determine the BH mass (e.g., \citealt{Peterson2004ApJ}).  However, the observations are too sparsely sampled to yield a robust measurement of the lag between the broad emission-line and continuum light curves, which, in any case, is dominated by an exponential ($t^{-5/3}$) decline. Instead, we estimate the virial BH mass from the single-epoch spectra, using the broad component of the Balmer lines. \cite{Ho2015ApJ} provide \hb-based virial mass estimators separately for classical and pseudo bulges.  In light of the properties of the host galaxy discussed in Section~\ref{sec:beforeout}, we adopt the zero point appropriate for pseudo bulges.  The highest resolution spectrum taken on 16 June 201 yields $M_\mathrm{BH} = 1.85_{-0.47}^{+0.5} \times 10^{7}\, M_\odot$, which is consistent with the value of $M_\mathrm{BH} \simeq 1.9 \times 10^{7}\, M_\odot$ reported by \citet{Trakhtenbrot2019ApJ}.  However, the full spectral series analyzed in Section~\ref{sec:specmodelafter} yields masses that span more than an order of magnitude, from  $M_\mathrm{BH} = 1.21 \times 10^{7}\, M_\odot$ to $32.6 \times 10^{7}\, M_\odot$ (Figure\,\ref{fig:virialMbh}).  The \ha-based method of \cite{Greene2005ApJ} produces a similarly large range of $M_\mathrm{BH} = 0.98 \times 10^{7}\, M_\odot$ to $26.3 \times 10^{7}\, M_\odot$, with a median value $7.51 \times 10^{7}\, M_\odot$.  The greatest discrepancy between the two tracers occurs at $t \approx 150-250$ days, owing to the large differences between the widths of the two lines during this period (Figure\,\ref{fig:vshift}b).  The rapid changes in $M_\mathrm{BH}$ are clearly unphysical, reflecting the dynamic evolution of the BLR that is not yet virialized (Section~\ref{sec:physpict}).

Absent a trustworthy BH mass based on the broad emission lines, we turn to the properties of the host galaxies for alternatives. The BH mass can be estimated from the $M_\mathrm{BH}-\sigma_\ast$ relation \citep{Gebhardt2000ApJ,Ferrarese2000ApJ}. Since no stellar velocity dispersion is available for the bulge of 1ES\,1927+654, we assume $\sigma_g \simeq \sigma_\ast$, which has been shown to be approximately valid for a large variety of AGNs (e.g., \citealp{Nelson1996ApJ,Greene2005ApJb,Ho2009bApJ,Kong2018ApJ}), where $\sigma_g$ is the velocity dispersion of the ionized gas derived from the narrow forbidden emission lines.  Using the three spectra with the highest spectral resolution taken on 6--9 March 2018, we find, after correction for instrumental resolution, ${\rm FWHM} = 217.2 \pm 44.1\, \rm\,km\,s^{-1}$ for the \oiii~$\lambda 5007$ line, or $\sigma_g = 92.2 \pm 18.7\, \rm\,km\,s^{-1}$. From the $M_\mathrm{BH}-\sigma_\ast$ relation for pseudo bulges of \cite{Saglia2016ApJ}, $M_\mathrm{BH} = 5.1_{-3.2}^{+9.3} \times 10^{6}\, M_\odot$ (cyan dashed line and shaded region in Figure\,\ref{fig:virialMbh}).  As a cross-check, we also estimate the BH mass from the stellar mass of the bulge.  The $M_\mathrm{BH} - M_\mathrm{bulge}$ relation given by \citet{Kormendy2013ARAA} only pertains to elliptical galaxies and classical bulges.  We derive the $M_\mathrm{BH} - M_\mathrm{bulge}$ relation for pseudo bulges using the sample presented in \citet{Kormendy2013ARAA}, fixing the slope to that of classical bulges (1.17) given by Kormendy \& Ho, and solving for the zero point and intrinsic scatter.  The relation for pseudo bulges is $\log{M_\mathrm{BH}} = 1.17 \log M_\mathrm{bulge}-4.61$, which has an intrinsic scatter of 0.33 dex.  From the image decomposition described in Section~\ref{sec:GALFIT}, 1ES\,1927+654 has $B/T \simeq 0.44$ in the $K_s$ band, which, in combination with a rest-frame color of $(B-V)_0 = 0.42$ mag (Section~\ref{sec:beforeout}), results in $M/L_K = 0.32$ (0.05 dex scatter; \citealt{Kormendy2013ARAA}), $M_\mathrm{bulge} = 1.57_{-0.35}^{+0.40} \times 10^{9}\, M_\odot$, and hence $M_\mathrm{BH} = 1.38_{-0.73}^{+1.57} \times 10^{6}\, M_\odot$ (black dashed line and shaded region in Figure\,\ref{fig:virialMbh}).  As a final consistency check, the total stellar mass of $M_\ast = 3.56_{-0.35}^{+0.38} \times 10^{9}\, M_\odot$ (Section~\ref{sec:beforeout}) implies $M_\mathrm{BH} = 1.08_{-0.81}^{+3.21} \times 10^{6}\, M_\odot$ (orange dashed line and shaded region in Figure\,\ref{fig:virialMbh}) according to the $M_\mathrm{BH} - M_\ast$ relation of late-type galaxies from \citet{Greene2020ARAA}.

In summary, as illustrated in Figure\,\ref{fig:virialMbh}, the BH mass estimated using traditional single-epoch virial estimators changes with time and significantly exceeds (by a factor of $\sim 10$) the masses derived from three independent methods based on the properties of the host galaxy.  A mass as large as $M_\mathrm{BH} \approx 10^{7}\, M_\odot$ would also be highly unexpected for the relatively low stellar mass of the host galaxy ($M_\ast = 3.56_{-0.35}^{+0.38} \times 10^{9}\, M_\odot$).  We suspect that the rapidly evolving BLR of the changing-look event in 1ES\,1927+654 is not yet virialized, and therefore its broad emission lines are not suitable for estimating the BH mass.  In the following, we adopt $M_\mathrm{BH} = 1.38_{-0.73}^{+1.57} \times 10^{6}\, M_\odot$, based on the $M_\mathrm{BH} -  M_\mathrm{bulge}$ relation, since it has the smallest scatter.

\section{Discussion}
\label{sec:discussion}

\subsection{Witnessing the Transformation of a True Type 2 into a Type 1 AGN}
\label{sec:type1s}

The origin of the BLR in quasars and AGNs remains elusive after decades of research. The broad-line emission may arise from a multi-phase medium in which material can condensate into clouds through thermal instability (\citealp{Krolik1981ApJ}).  Photoionization calculations suggest that, for a specific emission line, the clouds that have the maximum reprocessing efficiency only span a narrow range of density and distance from the central AGN (``locally optimally emitting clouds''; \citealp{Baldwin1995ApJ}). However, clouds cannot survive destruction by Rayleigh–Taylor and Kelvin–Helmholtz instabilities (e.g., \citealp{Proga2015ApJ}). Confinement by the pressure of an external magnetic field may be necessary \citep{Rees1989ApJ}, or perhaps by the pressure gradient developed on the photoionized surface layer of the gas due to the incident radiation force (\citealp{Baskin2014MNRAS, Netzer2020MNRAS}).

Prior to the optical outburst, 1ES\,1927+654 was considered to be a ``true'' type~2 AGN, on account of its rapid spectral variability and absence of obscuration in the X-rays \citep{Boller2003AA}, as well as the failure to detect hidden broad emission lines from either optical spectropolarimetry or near-IR spectroscopy \citep{Tran2011ApJ}. Thus, 1ES\,1927+654 joined the ranks of a small subset of type~2 AGNs suspected to lack a BLR intrinsically (e.g., \citealp{Tran2001ApJ,Tran2003ApJ,Panessa2002AA}). The intrinsic absence of broad lines is most commonly associated with AGNs of very low accretion rate \citet{Ho2008ARAA}, as a consequence of a luminosity limit below which either a BLR cannot survive because its compact size exceeds the maximum allowed velocity \citep{Laor2003ApJ} or there is insufficient column density produced by a radiation-driven wind to sustain a BLR (\citealt{Elitzur2009ApJ,Elitzur2014}). However, the BLR can also vanish in a minority of highly accreting AGNs (e.g., \citealt{Ho2012ApJ,Miniutti2013MNRAS}), on account of factors (e.g., detailed radial distribution of a radiation-driven wind and radiation efficiency) other than the luminosity that can affect the final BLR column density \citep{Elitzur2016MNRAS}.

\citet{Tran2011ApJ} suggested that 1ES\,1927+654 might be powered by a highly sub-Eddington ($\dot{M}/\dot{M_\mathrm{E}} \approx 0.001$), advection-dominated accretion flow.  The low accretion luminosity would lead to a very compact BLR, if it were to obey the $R_\mathrm{BLR}-L$ relation, whose broad lines would be either too broad and/or too faint to be detected. NGC\,3147 might serve as a possible analog \citep{Bianchi2019MNRAS}. However, 1ES\,1927+654 was bright in the X-rays even before the optical outburst. The 2011 XMM-Newton observation by \cite{Gallo2013MNRAS} indicated $L_\mathrm{2-10\,keV}=2.4 \times 10^{42}\rm\,erg\,s^{-1}$, which, with the luminosity-dependent X-ray bolometric correction of \citet{Duras2020AA}, yields a bolometric luminosity of $L_\mathrm{bol} \simeq 11.5\, L_\mathrm{2-10\,keV} = 2.7 \times 10^{43}\rm\,erg\,s^{-1}$. Assuming $M_\mathrm{BH} = 1.38 \times 10^{6}\, M_\odot$ (Section~\ref{sec:BHmass}), the corresponding Eddington ratio is $\lambda_\mathrm{E} \equiv L_\mathrm{bol}/L_\mathrm{E} = 0.16$, with the Eddington luminosity $L_\mathrm{E}=1.26 \times 10^{38}\,(M_\mathrm{BH}/M_\odot)$.  If we consider the $\lambda_\mathrm{E}$-dependent bolometric correction of \citet{Vasudevan2009MNRAS}, $L_\mathrm{bol} \simeq 70\, L_\mathrm{2-10\,keV} = 1.7 \times 10^{44}\rm\,erg\,s^{-1}$, and $\lambda_\mathrm{E} = 0.97$. In either case, it appears unlikely that 1ES\,1927+654 was ever faint enough to qualify as advection-dominated.  Instead, we suggest that the BLR was intrinsically absent before the optical outburst, possibly due to the low column density of the clouds launched by a radiation-driven disk wind.  In the disk-wind scenario of \citet{Elitzur2016MNRAS}, the minimal bolometric luminosity to ensure the appearance of broad lines must satisfy $L_\mathrm{min >} = 3.5\times 10^{44} \, (M_\mathrm{BH}/10^{7}\, M_\odot)^{2/3} \rm\,erg\,s^{-1}$.  For 1ES\,1927+654, this threshold is $L_\mathrm{min >}  = 0.93 \times 10^{44}\rm\,erg\,s^{-1}$, which, depending on the assumed X-ray bolometric correction, is comparable to or significantly higher than the pre-outburst bolometric luminosity. This may explain why 1ES\,1927+654 originally lacked broad lines.  The relatively high spin of the BH in 1ES\,1927+654 ($a_\ast \approx 0.8$, as inferred from broad-band SED 
fitting; \citealt{Li2022}) may be an additional contributing factor, as the likelihood of a source to be a true type~2 AGN significantly increases with increasing radiation efficiency \citep{Elitzur2016MNRAS}. After the outburst, the optical luminosity of 1ES\,1927+654 reached $L_{5100} = 2.30 \times 10^{43}\rm\,erg\,s^{-1}$, or $\lambda_\mathrm{E} \approx 1.3$ for $L_\mathrm{bol} = 10\,L_{5100}$ \citep{Richards2006ApJS}.  The X-ray luminosity also increased 10-fold \citep{Ricci2020ApJL}.  During the changing-look event, 1ES\,1927+654 lied comfortably above the threshold for BLR formation \citep{Elitzur2016MNRAS} and formally would be considered super-Eddington, consistent with its very steep X-ray power-law spectrum ($\Gamma \approx 3$; \citealp{Ricci2021ApJS}), which is usually associated with super-Eddington AGNs (e.g., \citealp{Trakhtenbrot2017MNRAS,Ricci2018MNRAS}).  

At the same time, 1ES\,1927+65 displays some striking differences compared to ordinary AGNs accreting at or near the Eddington limit.  Perhaps the most suitable comparison are the narrow-line Seyfert 1 galaxies \citep{Osterbrock1985ApJ}, a subclass of type~1 AGNs characterized by narrow \hb\ lines (${\rm FWHM} \lesssim 2000\, \rm km\,s^{-1}$) and strong \feii\ emission.  These properties are a consequence of their relatively low BH masses ($M_\mathrm{BH} \approx 10^{6}-10^{7}\, M_\odot$) and high Eddington ratios (\citealp{Boroson2002ApJ,Shen2014Nature}). By contrast, 1ES\,1927+65, notwithstanding its modest BH mass ($M_{\rm BH} \approx  10^{6}\, M_\odot$) and high Eddington ratio ($\lambda_{\rm E} \gtrsim 1$),  exhibits broad \hb\ emission with line widths that range from ${\rm FWHM} \approx 4000\, \rm km\,s^{-1}$ to nearly 15,000 $\rm km\,s^{-1}$.  This strongly suggests, as we already argued in connection with the unreasonable virial BH masses obtained in Section~\ref{sec:BHmass}, that the line-emitting region has yet to reach virialization.  Another notable distinction lies in the exceptional weakness of \feii\ emission found in the source.  The ratio\footnote{Following \citet{Boroson1992ApJS}, the equivalent width of \feii\ was calculated between 4434\angs\ and 4684\angs\ based on the best-fit model in Section~\ref{sec:specmodelafter}.} between the equivalent width of \feii\ and \hb, $R_{\rm Fe~II} = 0.16^{+0.13}_{-0.11}$, is much less than the value of $R_{\rm Fe~II} \approx 1$ expected for AGNs with $\lambda_\mathrm{E} > 1$ (\citealp{Boroson1992ApJS,Boroson2002ApJ,Dong2011ApJ,Shen2014Nature}). \citet{Boroson2002ApJ} argued that at high Eddington ratios, the presence of a soft X-ray excess would produce a large zone of warm, partially ionized, \feii-emitting gas. And yet, 1ES\,1927+654 emits conspicuously strong soft X-ray emission \citep{Ricci2021ApJS}, in strong conflict with its low $R_{\rm Fe~II}$.  

Interestingly, \feii\ emission also appears to be weak in TDEs reported in other AGNs suspected to be experiencing super-Eddington accretion (e.g., AT\,2018dyb; \citealp{Leloudas2019ApJ}).  \citet{Shields2010ApJ} argue that the strength of \feii\ emission in AGNs is driven not primary by $\lambda_\mathrm{E}$ but instead by gas-phase iron abundance.  This argument is buttressed by the recent photonionization calculations of of \citet{Panda2021}, who find that iron must be overabundant to reproduce correctly the observed values of $R_{\rm Fe~II}$. If, as we contend (Section~\ref{sec:physpict}), that the BLR clouds in 1ES\,1927+65 arose from the condensation of a wind that emanates from an accretion disk freshly created from the debris of a TDE, we naturally expect the material to bear the chemical imprint of the tidally disrupted star, which, in turn, should reflect the metallicity of the overall stellar population of the surrounding galaxy.  With a total stellar mass akin to that of the Large Magellanic Cloud (Section~\ref{sec:beforeout}), 1ES\,1927+65, too, should have a stellar metallicity of $\sim$1/2 solar \citep{Choudhury2016MNRAS}, which is substantially lower than the super-solar values normally inferred for the BLR (e.g.,  \citealp{Hamann1999ARAA,Jiang2007AJ}). The low metallicity of the clouds might naturally explain the non-detection of typical UV emission lines (e.g., \mgii) in the HST spectrum taken $\sim250$ days after the outburst \citep{Trakhtenbrot2019ApJ}. Meanwhile, the absence of resonance metal lines (e.g., \civ, \ciii) could reflect the temporary low ionization, which was argued to be the culprit for the lack of \heii\ then (Section~\ref{sec:heiilines}).

\begin{figure*}
\centering
\includegraphics[width=0.98\textwidth]{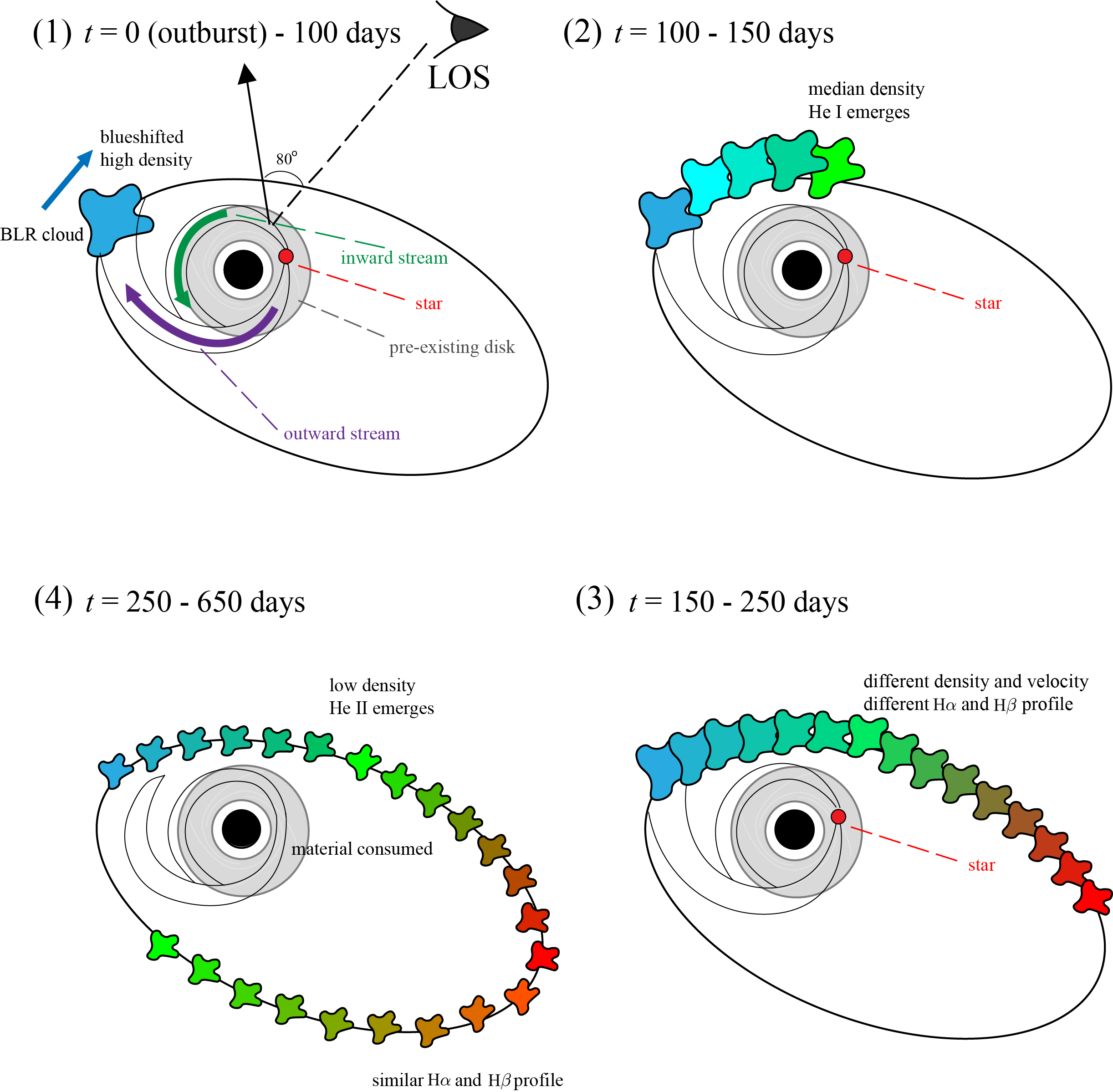}
\caption{Cartoon showing our proposed scenario for the formation and dynamical evolution of the BLR in the changing-look AGN 1ES\,1927+654.  Each colored cloud represents a collection of broad-line emitters that has similar location, radial velocity [blue for clouds moving toward our line-of-sight (LOS), red for clouds moving away], and density (larger cloud size indicates higher density). We schematically divide the evolution of the event into four successive stages. Stage (1): at $t = 0$ (outburst) to 100 days, a star (red dot) is tidally disrupted by the supermassive BH, forming a temporary accretion disk. The disk spreads both inward (green arrow) and outward (purple arrow).  The temporary, inward-moving accretion flow collides with the pre-existing accretion disk, while the material travelling outward moves on an eccentric Keplerian orbit (the outer ellipse). During this phase, BLR clouds can form above and below the elliptical accretion disk (with eccentricity $e \approx 0.6$; Section~\ref{sec:prochange}). The clouds initially have a non-zero radial velocity with respect to the observer (blue cloud). Stage (2): $t \approx 100-150$ days after the outburst, the outward-moving clouds have reached densities and an ionization level favorable for the production of \hei. Stage (3): at $t \approx 150-250$ days, the clouds have both positive and negative radial velocity with respect to the observer, occupy an even larger range in density and ionization, and produce different and asymmetric \ha\ and \hb\ profiles. Stage (4): by $t \approx 250-650$ days, the high accretion rate of the BH has consumed most of the material, resulting in a decrease of both the continuum and broad-line luminosity. Meanwhile, the low density of the inner clouds allows \heii\ to be excited, whose emergence coincides with the presence of weak, redshifted broad \ha\ emission.
}
\label{fig:cartoon}
\end{figure*}

\subsection{Physical Picture for the Formation of the BLR}
\label{sec:physpict}

The sudden appearance and rapid secular evolution of the BLR in the changing-look event in 1ES\,1927+654 offers us insights that, although not necessarily applicable to typical AGNs, nevertheless may help to elucidate the physical processes associated with the formation of broad emission lines.  From the evidence summarized in Section~\ref{sec:specevol}, we have shown that: (1) broad \ha\ and \hb\ were not detected until $t \approx 100$ days after the optical continuum reached its maximum, and $(\mathrm{H\alpha}/\mathrm{H\beta})_{b}$ increased with time; (2) \hei\ was detected at $t\approx150$ days, while it took $\sim 450$ days for \heii\ to appear; (3) both the continuum luminosity ($L_{5100} \propto t^{-5/3}$) and the FWHM of the broad Balmer lines decreased with time, plausibly suggesting that the changing-look event originated from a TDE \citep{Trakhtenbrot2019ApJ,Ricci2020ApJL,Ricci2021ApJS}; and (4) the BLR clouds are not virialized, but their velocity shifts are consistent with motions on an eccentric ($e \approx 0.6$) Keplerian orbit.

We propose that during the changing-look event in 1ES\,1927+654 we witnessed the formation of a BLR.  Before the outburst, the host galaxy was highly active but intrinsically lacked broad emission lines because its radiation-driven disk wind launched material with insufficient column density (Section~\ref{sec:type1s}).  The BLR was not formed until the overall luminosity increased above a critical value \citep{Elitzur2016MNRAS}, which was triggered by the TDE.  The picture of BLR formation we have in mind has many similarities to the model of \cite{Guillochon2014ApJ}, who used hydrodynamical simulations to investigate the dynamical evolution of the debris stream formed from the tidal disruption of a main-sequence star by a supermassive BH.  According to that work, the debris stream is confined by self-gravity in the two directions perpendicular to the original direction of the star, forming a transient accretion disk that spreads both inward and outward from the pericenter. The authors suggest that the debris stream has a negligible surface area and thus does not contribute significantly to the broad-line emission. Instead, the broad lines are produced by the clouds in the region above and below the forming accretion disk.  The model of \citet{Guillochon2014ApJ} can reproduce well our observations, and we use it to divide the formation and evolution of the BLR in 1ES\,1927+654 into four stages\footnote{Figure\,\ref{fig:opdetail} is designed to show the representative line profile of the four stages illustrated chronologically in Figure\,\ref{fig:cartoon}.}, as schematically sketched in Figure\,\ref{fig:cartoon}:

\begin{itemize}

\item{{\it Stage 1}: Around December 2017, an optical outburst occurred as a result of a sudden increase of the BH accretion rate induced by the tidal disruption of a star. Debris from the disrupted star formed a temporary accretion disk, which spread both inward and outward \citep{Guillochon2014ApJ}. The inward-moving accretion flow collided with the pre-existing accretion disk (panel 1), efficiently removing angular momentum and subsequently enhancing the BH accretion rate \citep{Chan2019ApJ}. The enhanced accretion rate depleted the inner disk, which led to the disappearance of the X-ray corona \citep{Ricci2020ApJL}.  Meanwhile, the accretion flow that travelled outward moved on an eccentric Keplerian orbit (panel 1). The orbit, described in detail in Section~\ref{sec:prochange}, is eccentric ($e\simeq 0.6$), and its orbital plane has an inclination angle of $\sim 80^\circ$ with respect to our LOS. The BLR clouds formed above and below the outer region of the eccentric accretion disk, via mechanisms such as condensation from a warm, radiation pressure-driven disk wind (e.g., \citealp{Elvis2017ApJ}). Adequate column density ($> 5\times 10^{21} \rm \, cm^{-2}$; \citealp{Netzer2013book}) was available after the optical outburst from the increase in accretion rate (\citealp{Elitzur2016MNRAS}; see also discussion in Section~\ref{sec:type1s}), allowing BLR clouds to form due to thermal instability in the radiatively heated and cooled medium \citep{Krolik1981ApJ}. For a typical AGN environment, the cloud formation timescale is $\sim 50$ days according to radiative hydrodynamical simulations of \cite{Proga2015ApJ}, while the acceleration timescale is $\sim 20$ days to reach the local sound speed. These timescales are somewhat shorter than the observed $\sim 100$-day lag between the emergence of the broad lines and the optical outburst, perhaps owing to the lower cooling efficiency induced by the low-metalicity environment of the host galaxy (Section~\ref{sec:type1s}).  Once formed, most of the clouds retain the original kinematics, which follow a Keplerian orbit that initially moves toward the LOS\footnote{As mentioned in Section~\ref{sec:heiilines}, $\sim 10\%$ of the clouds had outflow kinematics that depart from Keplerian motion.}.  The newly formed BLR clouds produce blueshifted, broad emission lines, as shown in the first row of Figure\,\ref{fig:opdetail}.}

\item{{\it Stage 2}: From 100 to 150 days after the ourburst, new BLR clouds continued to form, such that the broad-line luminosity increased without an accompanying increase in the continuum.  As the size of the accretion disk and the BLR expanded, the width of the broad Balmer lines dropped quickly (Figure~\ref{fig:vshift}b). Meanwhile, the gas number density ($n_\mathrm{H}$) also decreased \citep{Guillochon2014ApJ}, leading to the emergence of \hei. The ionization potential for \hei\ (24.6\,eV) being larger than that of \hi\ (13.6 eV), the strength of the \hei~$\lambda5877$ recombination line should be proportional to the ionization fraction when \hei\ is not fully ionized. Considering the balance between photoionization and recombination, the ionization fraction can be estimated as $n_\mathrm{X^+}/n_\mathrm{X^0}\propto I\mathrm{(X^0)}/n_{e}$, where $I\mathrm{(X^0)}$ is the incident rate of photons capable of ionizing the lower state ($\mathrm{X^0}$) to the higher state ($\mathrm{X^+}$), and $n_{e}$ is the electron number density. We suggest that initially the BLR clouds were too dense ($n_\mathrm{H} \approx 10^{12} \,\rm cm^{-3}$) for \hei\ to be ionized. \hei\ appeared only after the clouds became dilute enough ($n_\mathrm{H} \approx 10^{10}-10^{11} \,\rm cm^{-3}$; panel 2) for efficient ionization of \hei.}

\item{{\it Stage 3}: From 150 to 250 days, the broad Balmer lines reached their highest intensity as more and more BLR clouds formed.  The clouds occupied a wide range of spatial distribution surrounding the BH (panel 3). Not only did the gas number density ($n_\mathrm{H}$) decrease for the outlying clouds, the hydrogen ionizing flux [$\Phi(\mathrm{H})$] also dropped for the outward-moving clouds, since $\Phi(\mathrm{H}) \propto L\,r^{-2}$, such that as the event progressed the BLR clouds covered an increasingly larger space in the $\Phi_\mathrm{H}-n_\mathrm{H}$ plane. This scenario also explains the line profiles illustrated in the second and third rows of Figure\,\ref{fig:opdetail}: photoionization calculations (e.g., \citealp{Korista2004ApJ}) indicate that the equivalent width of broad \hb\ and \ha\ responds differently to $n_\mathrm{H}$ and $\Phi_\mathrm{H}$, wherein $\mathrm{H\alpha}/\mathrm{H\beta}$ increases with decreasing $n_\mathrm{H}$ and $\Phi_\mathrm{H}$.  The dependence of density on galactocentric distance (\citealp{Guillochon2014ApJ}) naturally predicts that the strength of broad \hb\ and \ha\ should vary with location along the elliptical orbit, providing a natural explanation for the observed variations in line profile and line ratio.}

\item{{\it Stage 4}: Beyond $\sim 250$ days, the high accretion rate of the BH has consumed most of the material (panel 4), resulting in a decrease of both the continuum and broad-line luminosity. Broad \hb\ and \ha\ began to fade systematically, and their profiles exhibited a symmetric, redshifted core because the lines were generated mostly by the outer clouds (fourth row of Figure\,\ref{fig:opdetail}).  The late emergence of \heii~$\lambda4686$ can be explained qualitatively in the same manner as \hei\ during stage 2.  With an ionization potential of 54\,eV, \heii\ is ionized significantly at this stage---indeed, almost fully ionized by the last epoch (Figure\,\ref{fig:heiiuplim}c)---because the inner clouds were rarefied to even lower density.}
\end{itemize}

The above physical model offers useful constraints on the life time of the clouds in 1ES\,1927+654. Clouds newly formed in a radiative environment cannot be sustained against destruction by Rayleigh–Taylor and Kelvin–Helmholtz instabilities. According to the hydrodynamical simulations of \cite{Proga2015ApJ}, isobaric clouds can dissolve as quickly as $t_\mathrm{life} \approx 10$ days, although denser clouds ($n_\mathrm{H} \ge 10^{12}\rm \, cm^{-3}$), with their sufficiently large cooling rate, potentially can be non-isobaric and stabilized by conduction \citep{Waters2019ApJ}. During Stage 3, the leading clouds evolve quickly, as evidenced by the notable differences of $\Delta V$, FWHM, and $S$ between broad \ha\ and \hb\ (Figure~\ref{fig:vshift}). After $\sim 250$ days, however, the differences disappear, while the line luminosities quickly drop (Figure~\ref{fig:hahb}). If we posit that these secular changes are induced by the fragmentation of the leading clouds, we can estimate $t_\mathrm{life} \approx 150$ days. Once fragmented, the clouds would span more evenly in space, possibly explaining the FWHM ``echo" in Figure~\ref{fig:vshift}b.

\citet{Scepi2021MNRAS}, followed by \citet{Laha2022arXiv}, suggest the alternative scenario that the changing-look event in 1ES\,1927+654 can arise from a change in the accretion rate and an inversion of magnetic flux in a magnetically arrested disk. In this picture, the peculiar optical and X-ray light curves result from an inward-leading advection event that brings magnetic flux of the opposite polarity.  This model, however, gives no predictions for the broad emission lines after the outburst event, and therefore no direct comparison can be made based on our spectroscopic analysis.  These authors attribute the non-detection of broad lines before the outburst to a low AGN power, which, however, is inconsistent with the high accretion rate we deduced for 1ES (Section~5.1).

\subsection{Comparison with other TDEs}
\label{sec:otherexp}

The evidence assembled in this study suggests that the BLR clouds of 1ES\,1927+654 condensed from a wind radiatively driven from a temporary accretion disk created by the TDE.  Here we draw some comparisons of this physical model with parallels seen in a few other TDEs.  In hydrodynamical simulation of TDE debris \citep{Guillochon2014ApJ}, the density evolution of the clouds is directly linked to their dynamical motion.  In the physical scenario offered in Section~\ref{sec:physpict}, \heii~$\lambda 4686$  emerges when the clouds have evolved to sufficiently low density to allow efficient photoionization of helium. Interestingly, the above scenario of \heii\  production implies that if the clouds initially were not fully ionized (below the blue dashed line in Figure~\ref{fig:heiiuplim}c), the \heii/\ha\  ratio would gradually increase with decreasing cloud density (see examples also in \citealp{Charalampopoulos2022AA}). TDEs that exhibit large \heii/\ha\  immediately after discovery (e.g., ASASSN-14li and ASASSN-14ae) might have clouds of intrinsically lower density or are helium-rich \citep{Gezari2012Nature}. 

The link between density evolution of the clouds and their dynamical motion offers an explanation as to why in 1ES\,1927+654 \heii~$\lambda 4686$ emerged significantly delayed (by $\sim 240$ days relative to the peak of the \ha\ emission) compared to the handful of other TDEs that initially had low \heii/\ha. In their detailed spectroscopic analysis of AT\;2018hyz, whose \ha\ and \hb\ response relative to the optical continuum closely resembles that of 1ES\,1927+654, \citet{Short2020MNRAS} report that \heii\ appeared roughly 70 days after the \ha\ peak, much quicker than the 240-day delay observed for 1ES\;1927+654.  To understand the difference, we note that the average FWHM of the \ha\ and \hb\ lines in AT\;2018hyz is $\sim 12,000\, \rm\,km\,s^{-1}$, about 1.5 times larger than in 1ES\,1927+654 (${\rm FWHM} \approx 8000\rm\,km\,s^{-1}$).  Given the similar BH mass ($M_\mathrm{BH} \approx 10^6\, M_\odot$; \citealp{Short2020MNRAS}), the broader emission lines of AT\;2018hyz imply that its BLR clouds are located at smaller semi-major radii, which have correspondingly shorter dynamical timescales: $\tau_{\rm dyn}^{\rm AT} = (a^{\rm AT}/a^{\rm 1ES})^{3/2} \tau_{\rm dyn}^{\rm 1ES} \simeq ({\rm FWHM}^{\rm AT}/{\rm FWHM}^{\rm 1ES})^{-3} \tau_{\rm dyn}^{\rm 1ES} \simeq 0.3 \tau_{\rm dyn}^{\rm 1ES}$.  If the emergence of \heii\ is related to the dynamical evolution of the disk, the factor of $\sim 3$ in dynamical timescale for the two TDEs can naturally account for the time difference in relative lag between \heii\ and \ha\ ($0.3\times240 \simeq 70$ days).  A similar argument can be made for the TDE PS18kh (\citealp{Holoien2019ApJ}). Although it has a larger BH mass ($M_\mathrm{BH} \approx 10^{6.9}\, M_\odot$) than 1ES\,1927+654, its \ha\ line is also broader (${\rm FWHM} = 12,000\,\rm\,km\,s^{-1}$), such that their dynamical timescales end up being roughly comparable: $\tau_\mathrm{dyn}^\mathrm{PS} \simeq (M_\mathrm{BH}^\mathrm{PS}/ M_\mathrm{BH}^\mathrm{1ES}) (\mathrm{FWHM}^\mathrm{PS}/ \mathrm{FWHM}^\mathrm{1ES})^{-3}\tau_\mathrm{dyn}^\mathrm{1ES} \simeq 1.6 \tau_\mathrm{dyn}^\mathrm{1ES}$. This may explain why the time delay for the emergence of \heii\ is almost identical for both, namely 250 days for PS18kh (\citealp{Holoien2019ApJ}) versus 240 days for 1ES\,1927+654.  The two TDEs bear a close resemblance in another intriguing aspect. As in 1ES\,1927+654 (see Section~\ref{sec:balmer_decrement}), the Balmer decrement of PS18kh increased over time, from $(\mathrm{H\alpha}/\mathrm{H\beta})_{b} \approx 3$ initially to $\sim 7$ at late times \citep{Holoien2019ApJ}.

The changing-look AGN SDSS\,J015957.64+003310.5 \citep{LaMassa2015ApJ,Zhang2021MNRAS} provides a final interesting case for comparison, for, like 1ES\,1927+654, its broad Balmer lines exhibit a long-term radial velocity shift (between 2000 and 2010) that may be interpreted as arising from the orbital motion of BLR clouds.  With $M_\mathrm{BH} \approx 10^{6}\, M_\odot$ and \ha\ ${\rm FWHM} \approx 6000\rm\,km\,s^{-1}$ \citep{Zhang2019MNRAS}, the dynamical timescale of SDSS\,J015957.64+003310.5 is close to twice that of 1ES\,1927+654, or $\sim 7$ years.  This is roughly comparable to the 10-year timescale over which velocity changes were detected.  SDSS\,J015957.64+003310.5 shares another important similarity with 1ES\,1927+654 in that its broad \hb\ line grossly overestimates by $\sim 2$ orders of magnitude the BH mass predicted by the $M_\mathrm{BH} - \sigma_\ast$ relation.

But not all the characteristics of 1ES\,1927+654 are reproduced in other TDEs. For example, the broad \ha\ and \hb\ of AT\;2018hyz showed no evidence of velocity shift, but instead were asymmetric and double-peaked during the first 100 days \citep{Short2020MNRAS}. Such double-peaked Balmer lines are found in TDEs (e.g., PTF09djl; \citealp{Arcavi2014ApJ}) whose profiles can be fit well with a relativistic, elliptical accretion disk model \citep{Liu2017MNRAS}.  If this model applies to AT\;2018hyz, it implies that its broad, double-peaked Balmer lines are produced in low-density, optically thin material just above the disk \citep{Gomez2020MNRAS}, conditions perhaps unfavorable for cloud condensation that would result in detectable changes in velocity or Balmer decrement. By contrast, the broad \ha\ and \hb\ lines of 1ES\;1927+654 remained quite symmetric throughout our observations ($S > 0.7$; Figure\,\ref{fig:hahb}b), decidedly distinct from the highly asymmetric or double-peaked broad profiles expected from a disk.  The disk model also has difficulty accounting for the velocity shifts of 1ES\;1927+654, which can be modeled by a simple Keplerian orbit (Figure\,\ref{fig:vshift}a). 

 The above-mentioned differences between 1ES\,1927+654 and other TDEs suggests an alternative explanation for the double-peaked Balmer line profile, namely that it might comprise two kinematic distinct components, a main Keplerian component and an additional outflow component, as commonly invoked in velocity-resolved reverberation mapping models of type~1 AGNs \citep{Pancoast2014MNRAS,Li2018ApJ,Williams2018ApJ}. If we were to view the TDE disk from a more face-on orientation, the outflowing broad-line clouds would contribute a blue horn to the original broad profile, but we would see little velocity shift evolution, as in the case of AT\;2018hyz. For a more edge-on view, the additional flux would be centered closer to the rest wavelength, and more significant velocity shift evolution would be seen, as in the case of 1ES\,1927+654 and ASASSN-18pg \citep{Holoien2020ApJ}. Since the two components can have intrinsically different light curves (Figure~\ref{fig:hahb}), the disappearance of the second \ha\ peak of AT\;2018hyz \citep{Short2020MNRAS} after $\sim100$ days can be attributed to the decrease of the flux of the outflow component. The two-component BLR model provides an alternative to the reprocessing scenario \citep{Dai2018ApJ} to explain the viewing angle dependence of the observed spectroscopic properties of TDEs \citep{Charalampopoulos2022AA}.

The roughly $t^{-5/3}$ decline of both the optical (Figure~\ref{fig:continuum}a) and UV \citep{Trakhtenbrot2019ApJ} continuum is consistent with the fallback accretion rate of TDEs \citep{Rees1988Nature}. While a similar light curve is expected in the X-rays \citep{Saxton2021SSRv}, this is not observed in 1ES\,1927+654 (\citealp{Ricci2020ApJL,Ricci2021ApJS}). \citet{Ricci2020ApJL} argue, based on the evolution of the X-ray hardness, that the X-ray corona might be destroyed around 200 days, after which it gradually gets recreated. Perhaps this event can be linked to the interaction between the pre-existing disk and the debris from a tidally disrupted star. In this respect, 1ES\,1927+654, already active prior to the outburst, fundamentally differs greatly from other TDEs.

\section{Summary}
\label{sec:summary}

1ES\,1927+654, historically a ``true'' type~2 AGN \citep{Boller2003AA, Tran2011ApJ} with relatively high accretion rate,  transitioned into a type~1 AGN on 23 December 2017 with the appearance of a blue continuum and broad optical emission lines \citep{Trakhtenbrot2019ApJ}.  The changing-look event was accompanied by a sharp drop and then rebound in the X-rays $\sim 200$ days after the outburst \citep{Ricci2020ApJL,Ricci2021ApJS}.  The optical continuum light curve and overall evolution of the broad emission lines suggest that the changing-look event was associated with a TDE.  In this work, we provide a detailed analysis of the properties of the host galaxy, with the primary aim of securing an independent constraint on the mass of the central BH.  To ascertain the host galaxy morphology, bulge type, and stellar mass, we perform two-dimensional image decomposition of optical images acquired with XMM-Newton prior to the outburst, supplemented by spectral energy distribution modeling of optical spectra and broad-band (UV, optical, and near-IR) photometry.  We study 34 optical spectra covering a period of $\sim 500$ days to conduct a thorough investigation of the spectral evolution of the AGN outburst, to elucidate the origin and characteristics of the BLR.

With a stellar mass of merely $M_\ast = 3.56_{-0.35}^{+0.38} \times 10^{9}\, M_\odot$, the host galaxy of 1ES\,1927+654 is relatively young (stellar population age $\simeq 1$ Gyr) and likely hosts a pseudo bulge that comprises 44\% of the total mass.  We use these mass measurements and an estimate of the stellar velocity dispersion of the bulge, in conjunction with standard scaling relations between BH mass and host galaxy properties, to bracket the BH mass in the range $\sim (1-5) \times 10^{6}\, M_\odot$, with a preferred, final adopted value of $M_\mathrm{BH} = 1.38_{-0.73}^{+1.57} \times 10^{6}\, M_\odot$.  At the peak of the AGN outburst, the bolometric luminosity of the BH exceeded the Eddington limit ($\lambda_\mathrm{E} \simeq 1.3$).

During the course of the monitoring campaign, 1ES\,1927+654 displayed dramatic spectral variability, as follows:

\begin{enumerate}

\item Contrary to the behavior in the X-rays, the optical continuum luminosity decreased monitonically with time, roughly as $t^{-5/3}$ as observed in TDEs.  The spectral index of the continuum was initially close to that predicted for a standard thin disk ($f_\nu \propto \nu^{+0.33}$), and then, after an erratic drop around 200 days after the outburst, gradually fell to a value typically found in normal quasars ($f_\nu \propto \nu^{-0.44}$).

\item The broad Balmer lines emerged at $\sim 100$ days, and their flux rose sharply to a prolonged maximum that extended from $\sim 120$ to 230 days. The Balmer decrement of the broad lines increased substantially and systematically from $(\mathrm{H\alpha}/\mathrm{H\beta})_{b} \approx 1$ to 5.

\item The kinematics of the broad Balmer lines changed in a complex but systematic manner.  The radial velocities were initially blueshifted, but they became redshifted by the end of the observation campaign, following a trend with time that can described by an eccentric ($e \approx 0.6$) Keplerian orbit.  The FWHM of broad \ha\ and \hb\ decreased by a factor $\sim 2$ from the time the lines first appeared to the last epoch of observation.  The profiles were roughly symmetric at the beginning and the end of the campaign, but they transitioned to a period of high asymmetry when they were strongest at $\sim 100-200$ days.

\item We detected a new component of narrow Balmer emission. Narrow \ha\ and \hb\ systematically increased in strength by as much as a factor of 4 after 200 days, and then gradually declined.  By the end of the last epoch they had not yet returned to their pre-outburst level.  The light curves for the narrow lines rose and fell more gradually than those of the broad lines.  

\item Helium lines were detected later than the hydrogen Balmer lines, at $\sim 150$ days for \hei~$\lambda5877$ and $\sim 450$ days for \heii~$\lambda4686$. The ratio of narrow \heii/\ha\ increased significantly at late times, until finally reaching the value for fully ionized nebular gas.

\end{enumerate}

The above observational characteristics paint a complex yet consistent physical picture for the changing-look event in 1ES\,1927+654.  We propose that the debris of a tidally disrupted star formed a temporary accretion disk.  The super-Eddington accretion event drove a strong disk wind, which promoted the condensation of clouds whose density and kinematics changed systematically throughout their orbital evolution, as manifested by the appearance of emission lines of varying strength, ionization, relative intensity, velocity width, and asymmetry. An unavoidable consequence of such a dynamically evolving BLR is that the common practice of using the broad lines to derive a virial mass for the central BH is invalid, as evidenced by the time-varying BH masses inferred from broad \ha\ and \hb, which appear grossly overestimated compared to the mass estimates based on the host galaxy properties. This study highlights the value of coordinated, long-term spectroscopic monitoring of TDEs and changing-look AGNs, which can serve as an effective laboratory for elucidating the physical processes around the central engine of supermassive BHs.

\acknowledgements
 We thank the anonymous referee for helpful suggestions. This work was supported by the National Science Foundation of China (11721303, 11991052, 12011540375) and the China Manned Space Project (CMS-CSST-2021-A04, CMS-CSST-2021-A06). CR acknowledges support from the Fondecyt Iniciacion grant 11190831. BT acknowledges support from the European Research Council (ERC) under the European Union's Horizon 2020 research and innovation program (grant agreement 950533) and from the Israel Science Foundation (grant 1849/19). IA is a CIFAR Azrieli Global Scholar in the Gravity and the Extreme Universe Program and acknowledges support from that program, from the European Research Council (ERC) under the European Union’s Horizon 2020 research and innovation program (grant agreement number 852097), from the Israel Science Foundation (grant number 2752/19), from the United States - Israel Binational Science Foundation (BSF), and from the Israeli Council for Higher Education Alon Fellowship.

\software{SAS (v18.0.0; \citealp{Gabriel2004ASPC}), SExtractor \citep{Bertin1996AAS}, GALFIT \citep{Peng2002AJ,Peng2010AJ}, photutils \citep{Bradley2019zndo}, IRAF \citep{Tody1986SPIE,Tody1993ASPC}, lmfit \citep{Newville2021zndo}, emcee \citep{ForemanMackey2013PASP}}


\appendix
\section{The Effect of Spectral Resolution on Velocity Shift Measurements}
\label{app:test}

Our interpretation of a rapidly evolving BLR relies on detailed decomposition of the multi-epoch spectra. However, the spectra from different instruments deliver a range of spectral resolution, from $R\approx 400$ for data taken with FLOYDS\footnote{\url{https://lco.global/observatory/instruments/floyds/}} to $R\approx 5000$ for spectra from DIS\footnote{\url{https://www.apo.nmsu.edu/arc35m/Instruments/DIS/}}. To test the robustness of our results, we generate a series of simulated spectra that resemble the spectral resolution and SNR of the data and repeat the spectral decomposition.

\begin{figure}
        \figurenum{A1}
	\centering
	\includegraphics[width=0.49\textwidth]{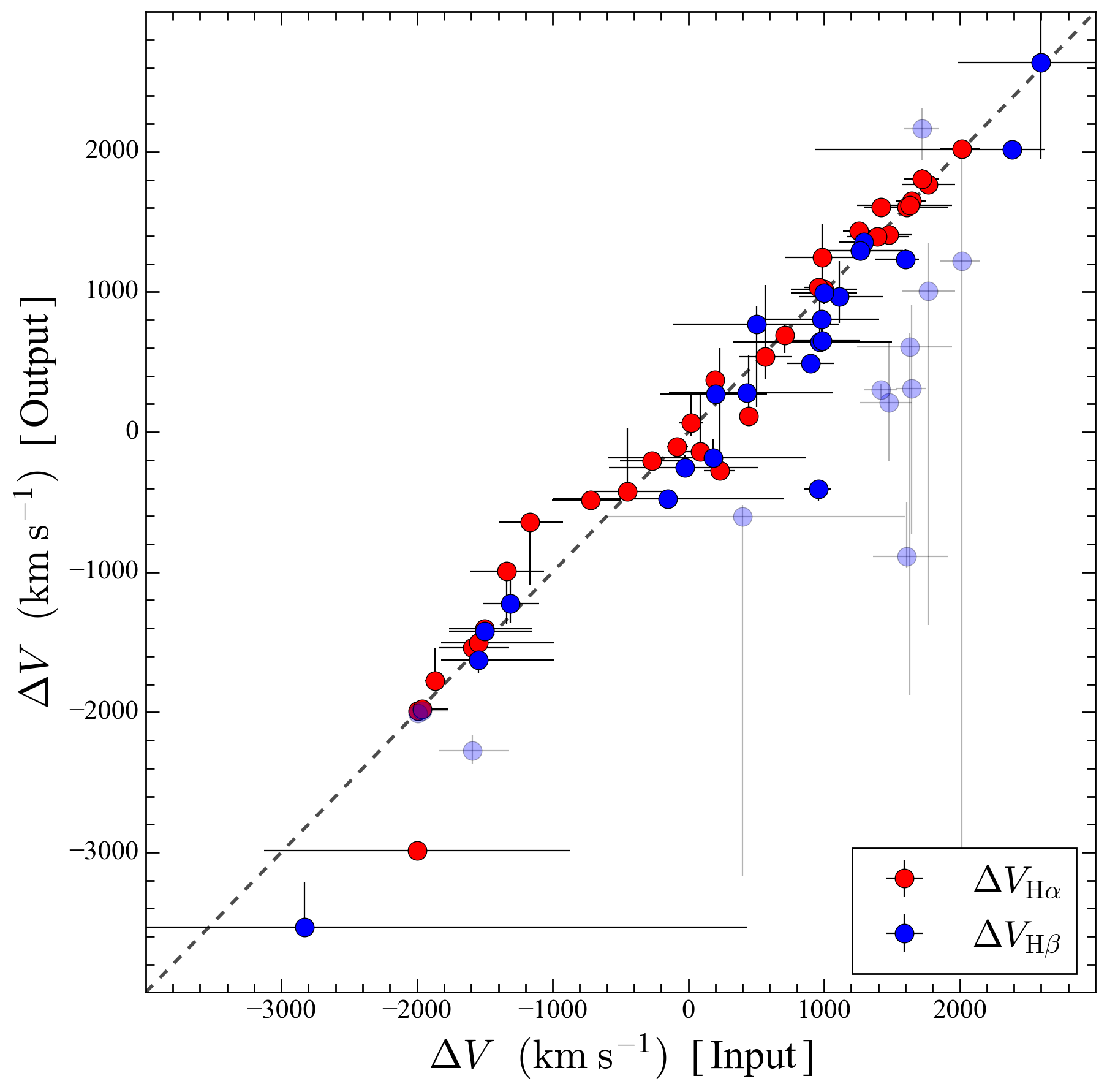}
	\caption{Simulation test on velocity shift ($\Delta V$) measurements for 1ES\,1927+654. The x-axis is derived from our MCMC best-fit model (Section~\ref{sec:specmodelafter}). Each point indicates an individual epoch.  The y-axis is derived from fitting simulated spectra; each point represents 100 realizations, where the error bar is calculated from the 16\% and 84\% values. Weak detections of \hb\ (SNR$\,<2$) are labeled by lighter color.  The dashed line denotes the one-to-one relation.
	}
	\label{fig:vstest}
	\end{figure}

We focus on the two spectral regions most relevant to our analysis, namely the \hb\ region ($4430-5200$~\AA) and the \ha\ region ($6100-7100$~\AA), and we mimic the conditions for the FLOYDS dataset, which has the sparsest sampling ($\Delta \lambda \approx 1.7$~\AA) and lowest spectral resolution ($R \approx 400$).  For each epoch of the observation, we simulate 100 mock spectra drawn from the SNR distribution, using the best-fit model of the original spectrum as the input. We perform least-squares minimization of the 100 realizations to derive the distribution of the best-fit parameters. Since the average FWHM of the broad Balmer lines ($\sim 9000\; \rm km\,s^{-1}$) is much larger than the lowest spectral resolution ($750\; \rm km\,s^{-1}$), the output values of the FWHM derived from the mock spectra are, as expected, quite consistent with our measurements.  The median differences are small for line width ($\Delta \rm FWHM/FWHM = 2\% \pm19 \%$) and broad Balmer line luminosity ($\Delta L_\mathrm{H\alpha}/L_\mathrm{H\alpha} = -2\% \pm14 \%$; $\Delta L_\mathrm{H\beta}/L_\mathrm{H\beta} = -0.5\% \pm18 \%$).  The output velocity shift, as illustrated in Figure\,\ref{fig:vstest}, shows good consistency for broad \ha\ $(\Delta V_\mathrm{H\alpha}^\mathrm{output}-\Delta V_\mathrm{H\alpha}^\mathrm{input})/\Delta V_\mathrm{H\alpha}^\mathrm{input} = -4\%\pm15 \%$).  The situation is much worse for broad \hb, which exhibits not only a large scatter but also a systematic bias\ $(\Delta V_\mathrm{H\beta}^\mathrm{output}-\Delta V_\mathrm{H\beta}^\mathrm{input})/\Delta V_\mathrm{H\beta}^\mathrm{input} = -25\% \pm41\%$) because of severe blending with broad \heii, especially when \hb\ is weak (SNR$\,<2$). Consequently, when fitting the evolution of $\Delta V$ with our Keplerian model (Section~\ref{sec:prochange}), we only use $\Delta V_\mathrm{H\alpha}$.

	\end{document}